\documentclass[aps,twocolumn,showpacs,groupedaddress, nofootinbib]{revtex4}  
\usepackage{graphicx} 
\usepackage{dcolumn}   
\usepackage{bm}        
\usepackage{amssymb}   
\usepackage{acronym}   
\usepackage{float}
\usepackage{xcolor}
\usepackage{hyperref}
\usepackage{latexsym}
\usepackage{url}
\usepackage{multirow}
\usepackage[flushleft]{threeparttable}
\usepackage{natbib}
\usepackage{subfigure}
\usepackage{amsmath}
\usepackage{caption}
\begin{document}

\widetext


\title{Binary Neutron Star Mergers and Third Generation Detectors: Localization and Early Warning}

\author{Man Leong Chan$^1$\footnote{m.chan.1@research.gla.ac.uk}, Chris Messenger$^1$, Ik Siong Heng$^1$ \& Martin Hendry$^1$}
\affiliation{$^1$SUPA, School of Physics and Astronomy, University of Glasgow, Glasgow G12 8QQ, UK}

\begin{abstract}
For third generation gravitational wave detectors, such as the Einstein 
Telescope, gravitational wave signals from binary neutron stars can last 
up to a few days before the neutron stars merge. To estimate the measurement 
uncertainties of key signal parameters, we develop a Fisher matrix approach 
which accounts for effects on such long duration signals of the time-dependent 
detector response and the earth’s rotation. 
We use this approach to characterize the sky localization 
uncertainty for gravitational waves from binary neutron stars at $40$, 
$200$, $400$, $800$ and $1600$Mpc, for the Einstein Telescope and Cosmic Explorer 
individually and operating as a network. We find that the Einstein Telescope alone 
can localize the majority of detectable binary neutron stars at a distance 
of $\leq200$Mpc to within $100\text{deg}^2$ with $90\%$ confidence. 
A network consisting of the Einstein Telescope and Cosmic Explorer can 
enhance the sky localization performance significantly -- with the $90\%$ credible region of $\mathcal{O}(1) \text{deg}^2$ 
for most sources at $\leq200$Mpc and $\leq100\text{deg}^2$ for most sources at $\leq1600$Mpc. 
We also investigate the prospects for third generation detectors identifying the presence of a signal 
prior to merger.
To do this, we require a signal to have a network signal-to-noise ratio of $\geq12$ and $\geq5.5$ for at least 
two interferometers, and to have a $90\%$ credible 
region for the sky localization that is no larger than $100 \text{deg}^2$. We find 
that the Einstein Telescope can send out such ``early-warning'' detection alerts 
$1$ - $20$ hours before merger for $100\%$ of detectable binary neutron stars 
at $40$Mpc and for $\sim58\%$ of sources at $200$Mpc. For sources at a distance of $400$Mpc, a 
network of the Einstein telescope and Cosmic Explorer can produce 
detection alerts up to $\sim3$ hours prior to merger for $98\%$ of detectable 
binary neutron stars.
\end{abstract}

\pacs{}
\maketitle

\acrodef{GW}[GW]{gravitational wave}
\acrodef{BNS}[BNS]{binary neutron star}
\acrodef{BBH}[BBH]{binary black hole}
\acrodef{NSBH}[NSBH]{neutron star black hole}
\acrodef{ET}[ET]{Einstein Telescope}
\acrodef{EM}[EM]{electromagnetic}
\acrodef{H}[H]{LIGO Hanford}
\acrodef{L}[L]{LIGO Livingston}
\acrodef{V}[V]{VIRGO}
\acrodef{SNR}[SNR]{signal to noise ratio}
\acrodef{MCMC}[MCMC]{Monte Carlo Markov Chain}
\acrodef{CE}[CE]{Cosmic Explorer}

\section{Introduction}
%
%
The first detections of \acp{GW} from \ac{BBH} systems GW150914, GW151226,
GW170104 and GW170608 by the two LIGO detectors at Hanford and Livingston
\cite{abbott2016observation, abbott2016gw151226, scientific2017gw170104, 2017ApJ...851L..35A} have
opened a new window on the universe and marked the beginning of \ac{GW}
astronomy.  In 2017, VIRGO began observation and the first joint detection
GW170814 was made by LIGO and VIRGO together \cite{abbott2017gw170814}.  Just a few days later, the three \ac{GW} observatories detected the first \ac{BNS}
merger event GW170817 \cite{PhysRevLett.119.161101}.  The detections of
multiple \ac{EM} counterparts associated with GW170817 initiated the era of \ac{GW} 
multi-messenger astronomy
\cite{abbott2017gravitational, 2017ApJ...848L..20M, goldstein2017ordinary,
abbott2017multi, tanvir2017emergence}.  In the coming years, 
additional ground-based interferometric detectors such as KAGRA
and LIGO India are likely to join the global network
\cite{aasi2016prospects, sathyaprakash1200219scientific,
aso2013interferometer}.  Many more \ac{GW}
detections can therefore be expected, from several different \ac{GW} sources in the universe such as compact
binary mergers (i.e. \ac{BBH}, \ac{BNS} and \ac{NSBH}), core-collapse supernovae, non-symmetric neutron stars and
the stochastic background \cite{abbott2017all, damour2005gravitational}.

%
For systems such as \acp{BNS} and \acp{NSBH}, 
the presence of a neutron star component is expected to lead to the generation
of associated \ac{EM} emission
accompanying the mergers of these systems -- including gamma ray bursts, x-ray
emission, kilonovae in the optical and infrared
bands and radio afterglows
\cite{cowperthwaite2015comprehensive, metzger2012most, tanvir2013kilonova,
berger2013r}.  Detecting an \ac{EM} counterpart in coincidence with a \ac{GW} trigger will increase the
detection confidence and improve the sky
localization of the \ac{GW} detection.  In addition, a successful \ac{EM}
follow-up observation can establish an association between the \ac{GW} trigger 
and its progenitor and provide a better understanding of the progenitor and its local
environment\cite{chassande2011multimessenger, 2009astro2010s20B,
kanner2008looc}.  A coincident detection of an \ac{EM} 
counterpart may also allow measurement of the redshift  of the source independently of the
\ac{GW} signal; this can be used for cosmological tests
and for constraining the equation of state of dark
energy \cite{schutz1986determining, sathyaprakash2010cosmography,
2017arXiv171005835A}.
Despite the low probability that
\ac{BBH} mergers will be accompanied by \ac{EM}
emission, identifying the host galaxy of a \ac{BBH} will still hopefully
establish the association between the \ac{BBH} and the galaxy.

%
One key factor to the success of \ac{EM} follow-up observations triggered by
\ac{GW} events, or that of identifying the host galaxy, is the
localization of the \ac{GW} source. However, \ac{GW}
interferometric detectors are not directional
instruments, and usually triangulation from a network of \ac{GW} detectors is
the approach to localizing \ac{GW} sources.  Therefore, their performances are
usually relatively poor when it comes to pinpointing sources of
\acp{GW}.  For example, the $90\%$ credible region for the sky positions
of the first detected BBH events are $\mathcal{O}(10^2)$ $\text{deg}^2$
\cite{abbott2016observation, abbott2016gw151226}.
The situation improved with VIRGO joining the observing runs, and it will
continue to improve when KAGRA and LIGO India start operating in later years
\cite{fairhurst2011source, nissanke2011localizing, nissanke2013identifying,
veitch2012estimating, rodriguez2014basic, sathyaprakash1200219scientific,
aasi2016prospects}.  However, for sources at large
distances, \ac{EM} follow-up observations will remain a difficult challenge. 
This is because a given type of \ac{GW} source observed at a larger distance will also have a larger associated localization error on the sky.  Larger distances also mean that the telescopes performing \ac{EM} 
follow-up observations will need to observe for longer to reach the sensitivity required to detect
the counterparts \cite{nissanke2013identifying}. As a result, the telescopes
may be unable to scan the whole $90\%$ credible region before the \ac{EM}
counterpart fades below the detectable threshold.  For example, the brightness
of kilonovae may peak at $1$ day after the merger and then start fading
\cite{metzger2012most} as was the case for the kilonova associated
with GW170817 \cite{tanvir2017emergence}.

%
Third generation detectors such as the \ac{ET} and
\ac{CE} can bring significant improvements in \ac{GW} source localization. 
Third generation detectors with enhanced sensitivity across the frequency band accessible to ground-based
detectors will be able to detect \acp{GW} from \ac{BBH} and
\ac{BNS} sources located at distances far beyond the horizon of second generation
detectors.  In particular, the improved sensitivity from $1$Hz to $10$Hz of
these detectors will distinguish them from second generation detectors, as this
allows for an extended duration of in-band observation of the signal.  Depending on the distance, signals from \ac{BNS} can be traced back
up to hours or days before merger. 

%
The long in-band duration of a signal will introduce several effects.
Firstly, it allows the detector to accumulate \ac{SNR}
over a significantly longer period of time.  As mentioned, one of the
difficulties in \ac{EM} follow-up observations of \ac{BNS} merger triggers at
large distance will be the time available to scan the $90\%$ credible region
associated with the \ac{BNS} trigger before the \ac{EM} counterpart becomes too
faint.  If the \ac{SNR} of a \ac{BNS} trigger can be accumulated to a
statistically significant level {\em prior\/} to the merger, prompt detection alerts could be made possible -- thus increasing the probability of detecting the \ac{EM} signature of the trigger \cite{cannon2011toward}.  The long duration also enables the detector to observe the source from different positions and directions as the earth rotates. This effect
is important in localizing the source of \acp{GW} as it result in the
time-dependency of the antenna pattern.  The \ac{GW} will also be Doppler shifted as the detector moves relative to the source as it rotates with the Earth's spin. The long duration of a signal thus requires the consideration of the earth's rotation when estimating the localization error.

%
In the literature, most studies of localization errors of \acp{GW} from
compact binary coalescences are based on the
assumption that the duration of the signal is short enough
that the rotation of the earth is negligible \cite{aasi2016prospects,
fairhurst2011source, chen2017facilitating}.  This is justifiable for the reason that the in-band durations of the signals in those
studies are only seconds to minutes in length.  
Mills, et al, 2017 \cite{2017arXiv170800806M} consider 
localization of short transient signals from \ac{BNS} mergers 
with a network of both second and three generation detectors.
More recently, there has
been work considering the long in-band duration and the rotation of the earth
\cite{zhao2017localization}. 
In this latter work the authors have modeled the \ac{GW} signal using the stationary phase approximation -- essentially the leading-order term in an expansion in powers of the small quantity that is the ratio of the radiation-reaction time scale and the orbital period.

In this work, we take into account the earth's rotation and, using a Fisher matrix approach, estimate the localization of \acp{GW} from \ac{BNS} sources observed by the \ac{ET} and \ac{CE} individually and as a
network.  We perform a series
of tests to estimate the localization capabilities of these detectors for
\acp{BNS} at $40$, $200$, $400$, $800$ and $1600$Mpc and for
a population of \ac{BNS} sources that are distributed uniformly in the comoving volume. We then investigate the feasibility of ``early-warning'' detection by setting requirements on localization error and accumulated \ac{SNR} before merger for an alert to be released.
Throughout this work, we focus our analysis on \acp{BNS} where the in-band duration of their signals can be days' long, and thus the
effect of earth's rotation is important. 

%
This paper is structured as follows: in Section\ref{sec:ET}, we briefly
introduce the configuration of the detectors we
considered for this work, and the
technologies employed to achieve their sensitivities.  The methodology is
presented in Section\ref{sec:method} with the results and simulations
presented in Section \ref{sec:simre} 
together with a discussion.  We then provide our conclusions
in Section \ref{sec:con}.

\section{Third Generation Detectors}\label{sec:ET}
%
%
There are currently two proposed  third
generation detectors: the \ac{ET} and \ac{CE}. For
the \ac{ET}, we employed the geometrical configuration known as ET-D as
discussed in \cite{sathyaprakash2012scientific} where the detector consists
of 3 individual interferometers. Each interferometer has an
opening angle equal to $60^\circ$ and is rotated by $120^{\circ}$ relative to the others, thus forming an equilateral triangle. The lengths of the arms of the interferometers are $10$km. 
The current design for \ac{CE} is an interferometric detector with a
$90^{\circ}$ opening angle and the arms as long as
$40$km \cite{LIGOW}.

\subsection{Location and Antenna Pattern}
%
%
The exact geographic locations at which the detectors will be built are
still unknown.  In this work, the location adopted for the \ac{ET} is
(Longitude, Latitude) = $(10.4^{\circ},~43.7^{\circ})$, and for \ac{CE}
is $(-119.41^{\circ},~46.45^{\circ})$.  Due to the differences between the \ac{ET} and \ac{CE} configurations, 
the antenna patterns of these detectors will be computed differently.  The antenna pattern of each
interferometer of the \ac{ET} can be expressed as \cite{regimbau2012mock}
\begin{eqnarray}\label{eq:inap}
F_+^1(\theta, \phi, \psi) =& -\frac{\sqrt{3}}{4}[(1+\cos^2\theta)\sin2\phi \cos2\psi \\ \nonumber
&+2\cos\theta \cos2\phi \sin2\psi], \\ \nonumber
F_{\times}^1(\theta, \phi, \psi) =& \frac{\sqrt{3}}{4}[(1+\cos^2\theta)\sin2\phi \sin2\psi \\ \nonumber
&- 2\cos\theta \cos2\phi \cos2\psi],  
\end{eqnarray}
\begin{eqnarray}\label{eq:23inap}
F_{+,\times}^2(\theta, \phi, \psi) =& F_{+,\times}^1(\theta, \phi + \frac{2\pi}{3}, \psi), \\ \nonumber
F_{+,\times}^3(\theta, \phi, \psi) =& F_{+,\times}^1(\theta, \phi - \frac{2\pi}{3}, \psi),
\end{eqnarray}
where the above equations assume that the detector is at the center of a spherical coordinate system in which $\theta$ and $\phi$ are the azimuthal angle and polar angle of the source respectively, 
and $\psi$ is the \ac{GW} polarization angle. The superscripts indicate the 3 interferometers in the configuration. 

A single interferometer with arms separated by $60^\circ$
will have a response
smaller by a factor of $\frac{\sqrt{3}}{2}$ compared to an $90^{\circ}$
interferometer.  However, the sum of three $60^\circ$ separated
interferometers will enhance the response by a factor of
$\sqrt{3}$, giving an overall factor of $\frac{3}{2}$ larger
than a single $90^{\circ}$ interferometric detector
configuration\cite{regimbau2012mock}.  The \ac{CE}'s
antenna pattern can be written as
\begin{eqnarray}\label{eq:tiap}
F_+(\theta, \phi, \psi) =& \frac{1}{2}(1 + \cos^2\theta)\cos2\phi\cos2\psi \\ \nonumber
&- \cos\theta\sin2\phi\sin2\psi, \\ \nonumber
F_\times(\theta, \phi, \psi) =& -\frac{1}{2}(1 + \cos^2\theta)\cos2\phi\sin2\psi\\ \nonumber
&-\cos\theta\sin2\phi\cos2\psi.
\end{eqnarray}

\subsection{Duration of Binary Neutron Star Signals}
%
%
In a \ac{GW} detector, different noise sources limit the sensitivity of the
detector in different frequency bands \cite{punturo2010einstein}. For
example, from $1$Hz to $10$Hz, seismic noise and gravity
gradient noise are the main contributors. From $10$Hz to $200$Hz, the
performance is limited by
quantum noise and thermal noise from the suspension and mirror coatings.  For frequencies
beyond $200$Hz, shot noise is the primary source of
noise \cite{2011LRR....14....5P}.

\begin{figure} 
\includegraphics[width=0.5\textwidth]{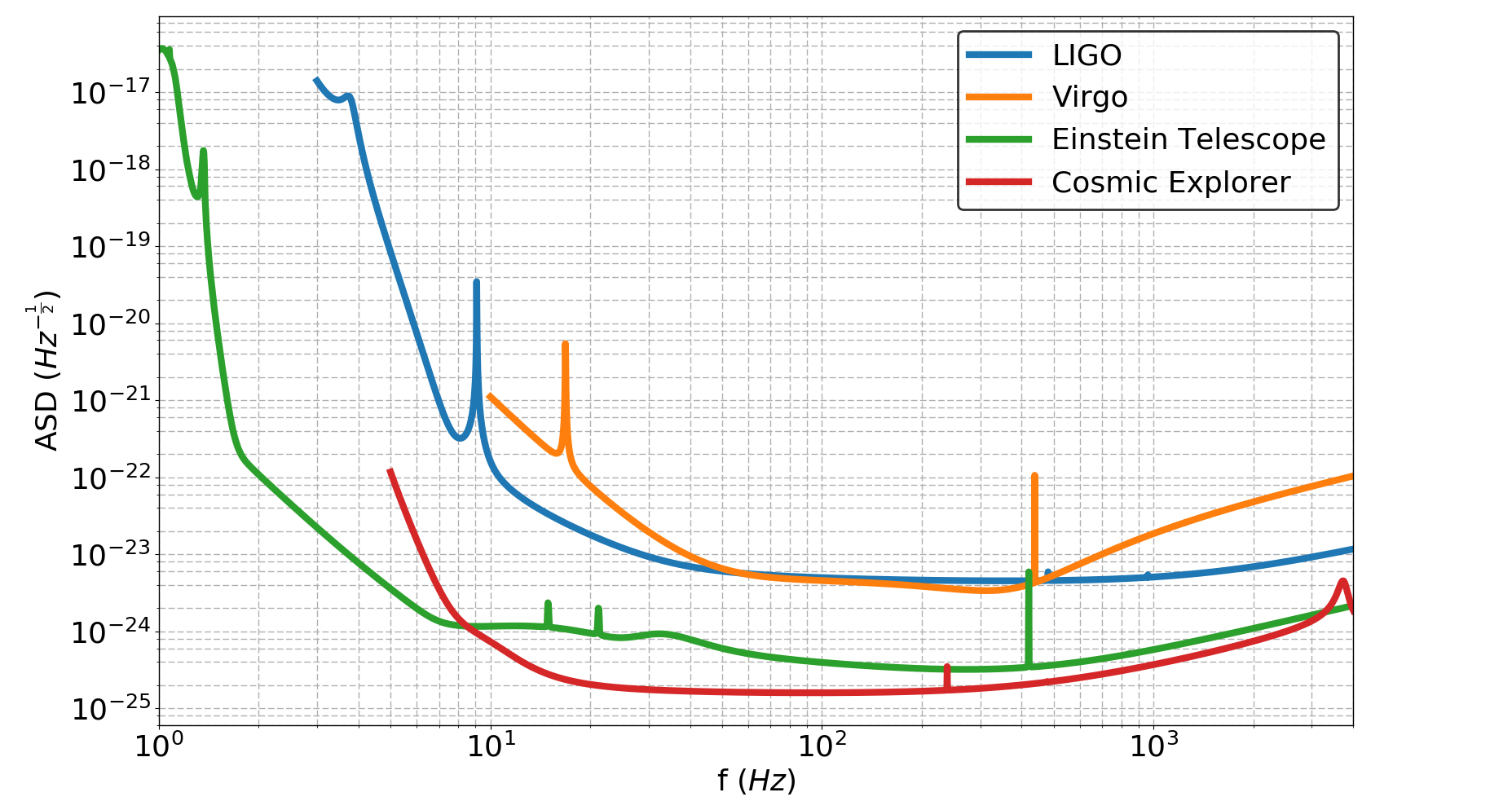}
\caption{The amplitude spectrum density for the \ac{ET} (Green), 
\ac{CE} (Red), aLIGO Hanford/Livingston (blue) and Advanced Virgo (Orange).  Both aLIGO and Advanced Virgo are at their respective design sensitivities.\label{fig:nos}} 
\end{figure}

%
There are many proposed technologies designed to improve the
sensitivity of the \ac{ET} \cite{hild2011sensitivity, hild2009xylophone, hild2008pushing}.  
The \ac{ET} is expected to have improved sensitivity across the frequency band accessible to ground-based detectors compared to 
second generation detectors.  
For \ac{CE}, the exact technologies that will be employed are still undergoing
research and study. The current sensitivity curve of \ac{CE} is obtained using
the existing technologies and well defined extrapolations from
them. As \ac{CE} will make use of the available technologies at the time that
it is built, the sensitivity curve should not be considered the design target
of \ac{CE} \cite{LIGOW}. The estimated amplitude spectrum density of the
\ac{ET} \cite{hild2008pushing} and \ac{CE} are presented in Figure
\ref{fig:nos}. The design sensitivity of aLIGO and Advanced VIRGO are also
shown for comparison. Compared to aLIGO and Advanced Virgo, the \ac{ET}'s
sensitivity from $10$Hz is improved by at least a factor of $10$, the
improvement can even be up to a factor of $10^6$ in the frequency band below
$10$Hz. \ac{CE}'s sensitivity is also improved by up to a factor of $10^3$ from
$1$Hz to $10$Hz compared to second generation detectors, and is improved
by $\sim30$ times above $10$Hz.

\begin{figure}
\includegraphics[width=0.5\textwidth]{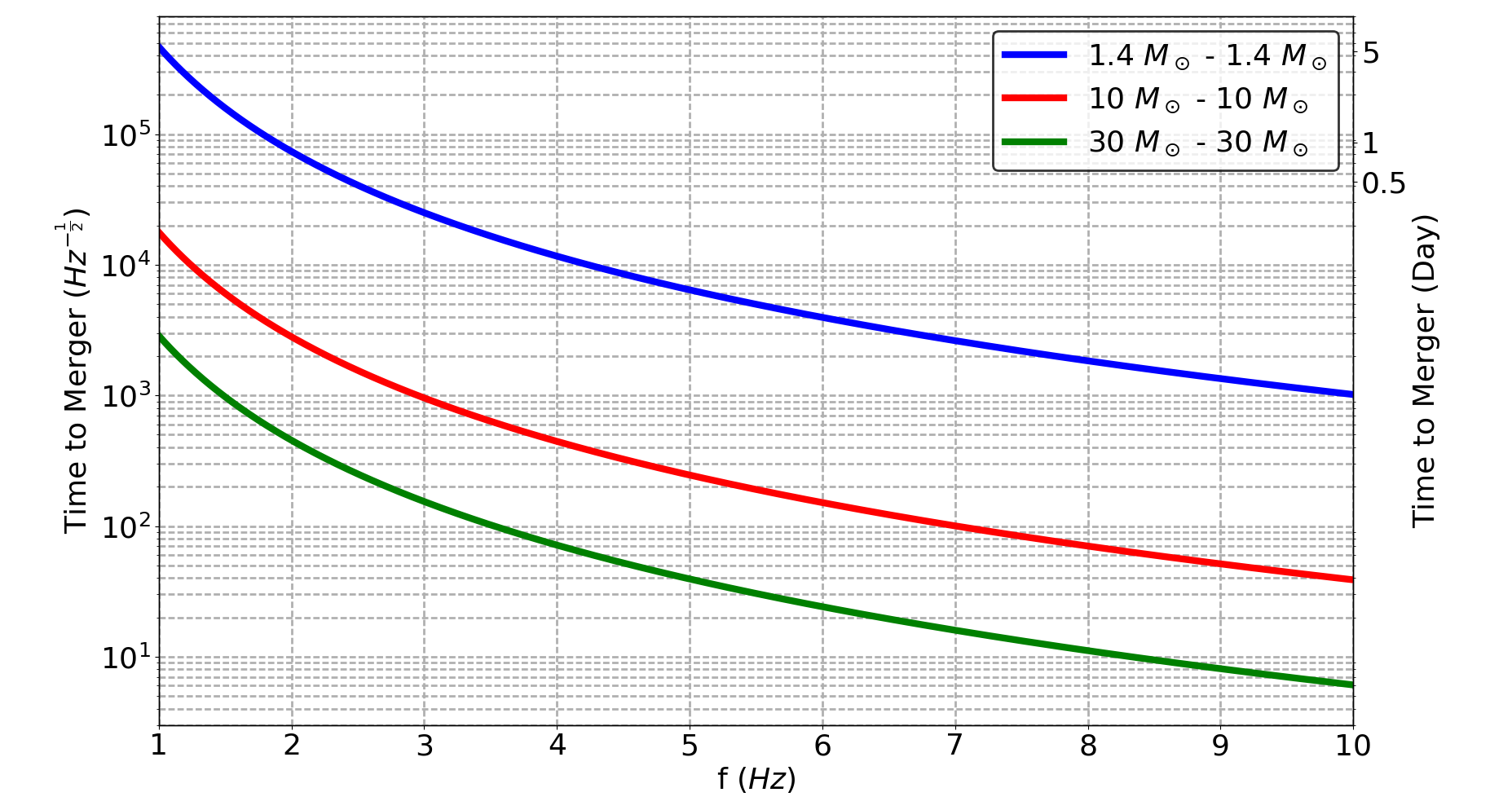}
\caption{The time to merger as a function of starting frequency $f_{\text{s}}$ for $1.4
M_{\odot} - 1.4 M_{\odot}$ \ac{BNS} (blue), $10 M_{\odot} - 10 M_{\odot}$ 
\ac{BBH} (red) and $30 M_{\odot}- 30 M_{\odot}$ \ac{BBH} (green). 
\label{fig:ToC}}
\end{figure}

%
Different relative improvements in sensitivity as a function of frequency 
can lead to different impacts on the sky localization capability. Better
sensitivity in the medium to high frequency band can effectively increase the
\ac{SNR} for a \ac{GW} event and thus reduce the 
localization error.  On the other hand, improvement in the low frequency band
might not increase the \ac{SNR} as much, but it will substantially extend the
in-band duration of the signal from the order of seconds/minutes to
the order of hours/days. We can express the time remaining prior to merger for a compact binary system as
a function of the instantaneous \ac{GW} frequency\cite{bassan2014advanced}
using
\begin{equation}\label{eq:ep1} 
\tau_\textrm{c} = \frac{5}{256}\frac{c^5}{G^{\frac{5}{3}}}\frac{(\pi f_{\text{s}})^{-\frac{8}{3}}}{\mathcal{M}^{\frac{5}{3}}},
\end{equation} 
where $\tau_\textrm{c}$ is the time to merger, $c$ the speed of light, $G$ the gravitational constant, $\mathcal{M}$
the chirp mass and $f_{\text{s}}$ the starting
frequency considered for the \ac{GW}. Figure
\ref{fig:ToC} shows the time to merger for a
$1.4M_{\odot} - 1.4M_{\odot}$ merger as a function of $f_{\text{s}}$.  Also plotted are $10M_{\odot} - 10M_{\odot}$ and $30M_{\odot}
- 30M_{\odot}$ \ac{BBH} mergers for comparison.  The in-band duration of a \ac{GW} from a given compact binary system in a detector can be obtained by replacing the
starting frequency $f_{\text{s}}$ with the low frequency cut-off 
of the detector in Eq.\ref{eq:ep1}. 
As indicated in Figure \ref{fig:ToC}, for \ac{BNS} systems, if
the detector's low frequency cut-off is reduced
to $2\text{Hz}$, the in-band duration of the signal will be close to $1$ day,
and will be more than $5$ days if the low frequency cut-off is $1\text{Hz}$.
This is substantially longer than the in-band duration for aLIGO and Advanced
VIRGO, where the low frequency cut-off is $10$Hz. 
The in-band duration of \ac{BBH} signals are expected to be shorter. 
For the \ac{ET} with low cut-off frequency at $1$Hz,  
$10M_{\odot} - 10M_{\odot}$ \ac{BBH} signals will last for $\sim 5$ hours.

\begin{figure*}
     \begin{center}
        \subfigure[]{%
            \label{fig:ET_1}
            \includegraphics[width=0.5\textwidth]{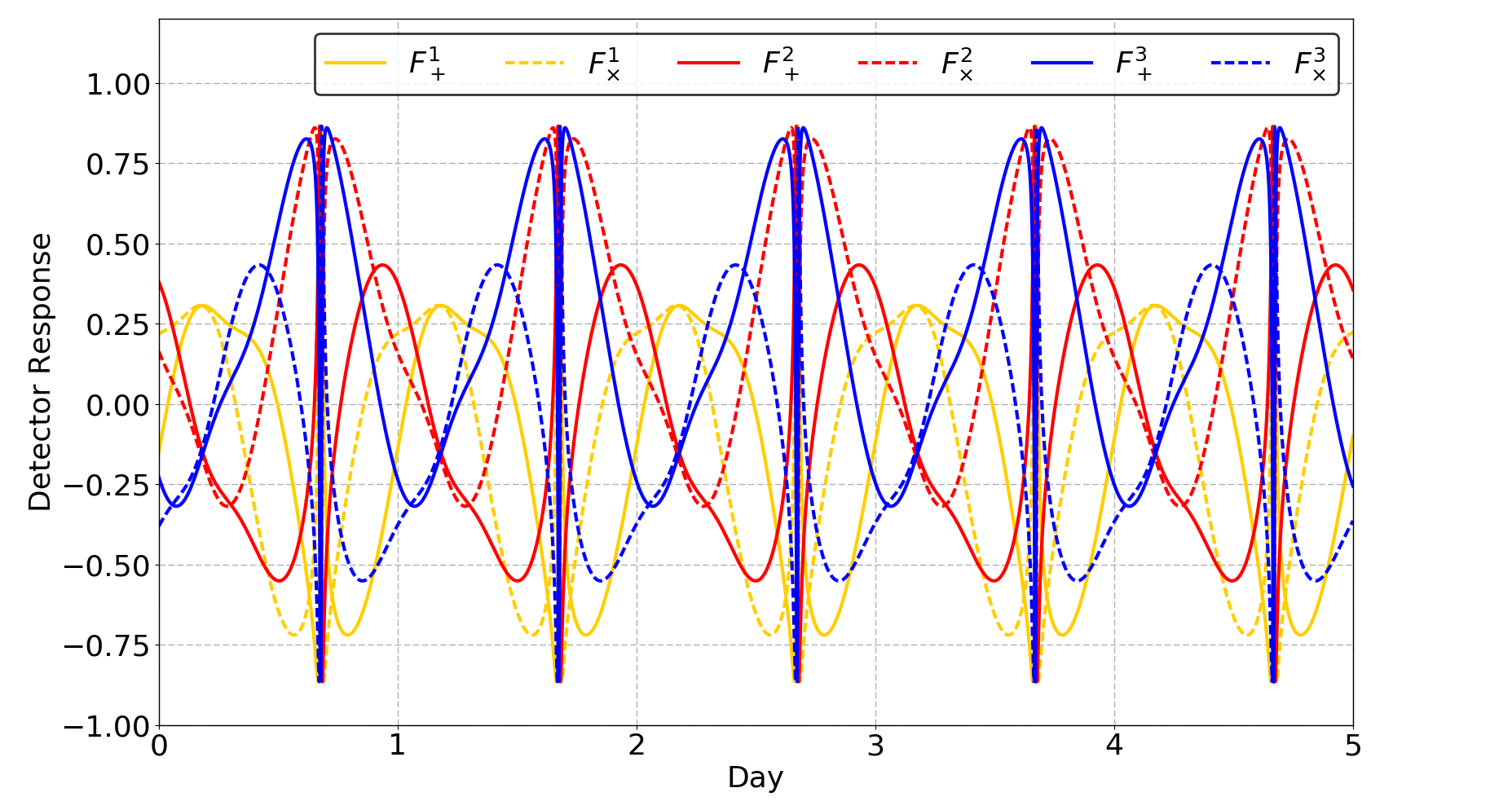}
        }%
        \subfigure[]{%
            \label{fig:ET_2}
            \includegraphics[width=0.5\textwidth]{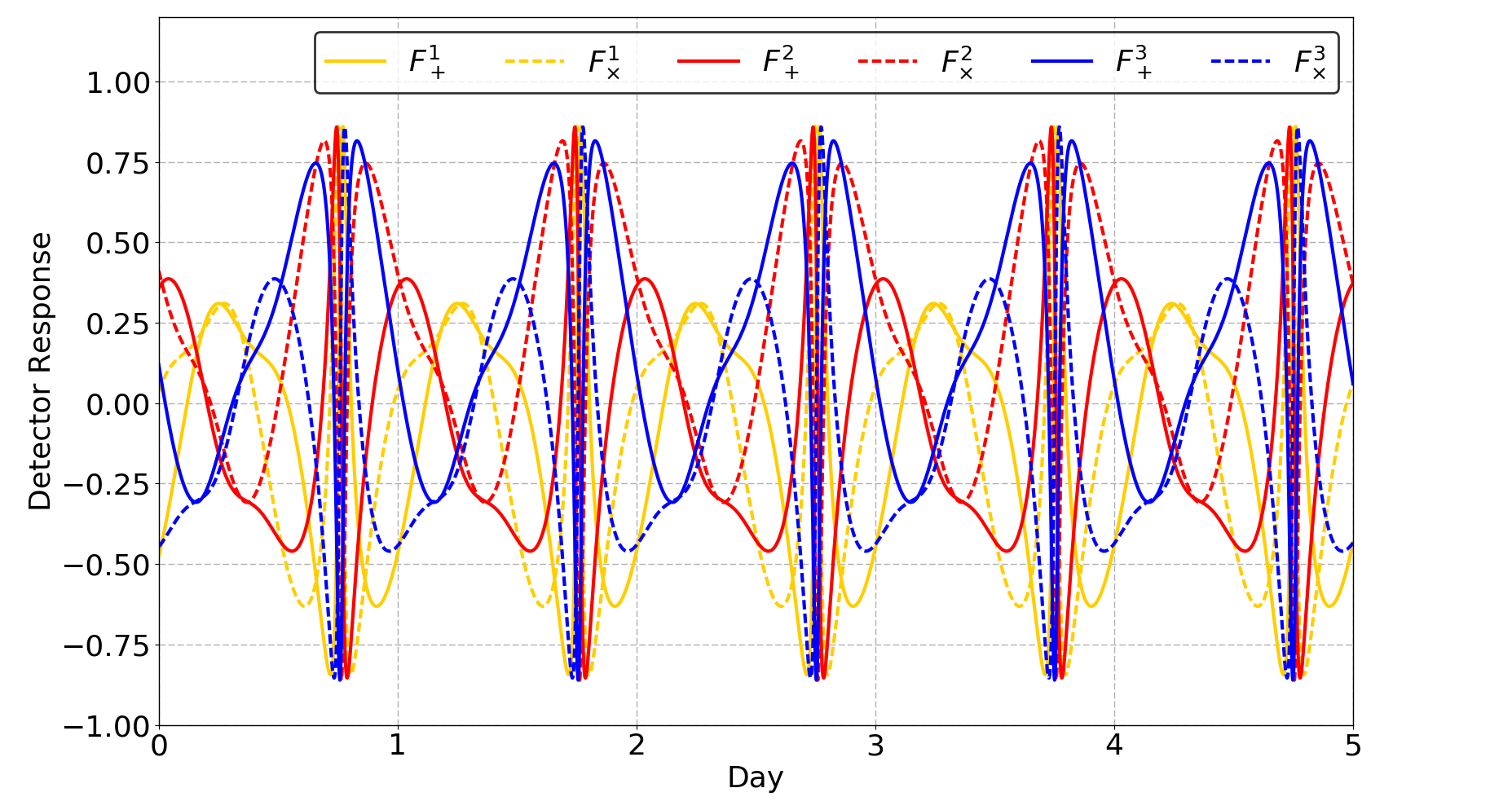}
        }%
\\
	\subfigure[]{%
            \label{fig:CE_1}
            \includegraphics[width=0.5\textwidth]{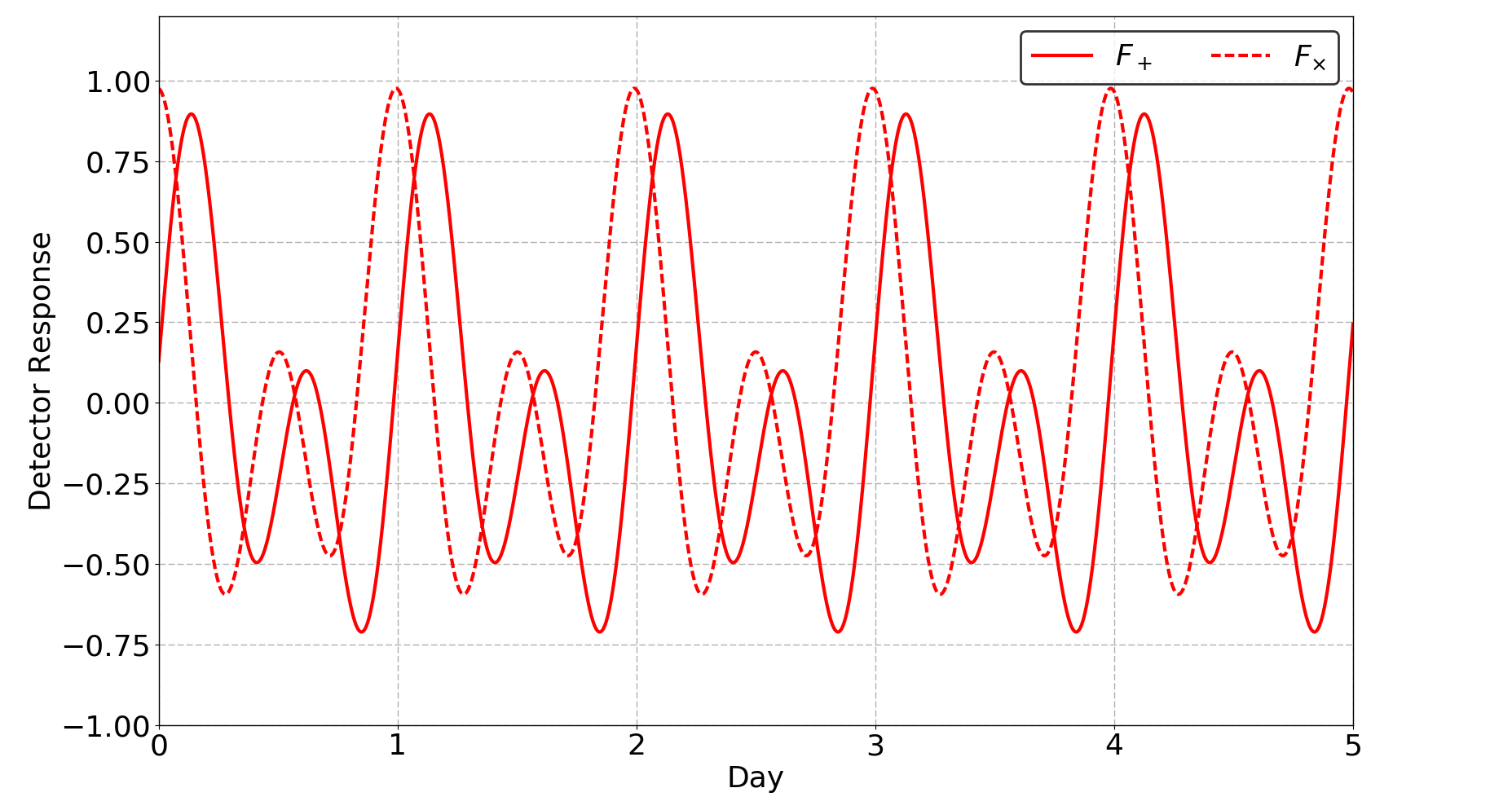}
        }%
        \subfigure[]{%
            \label{fig:CE_2}
            \includegraphics[width=0.5\textwidth]{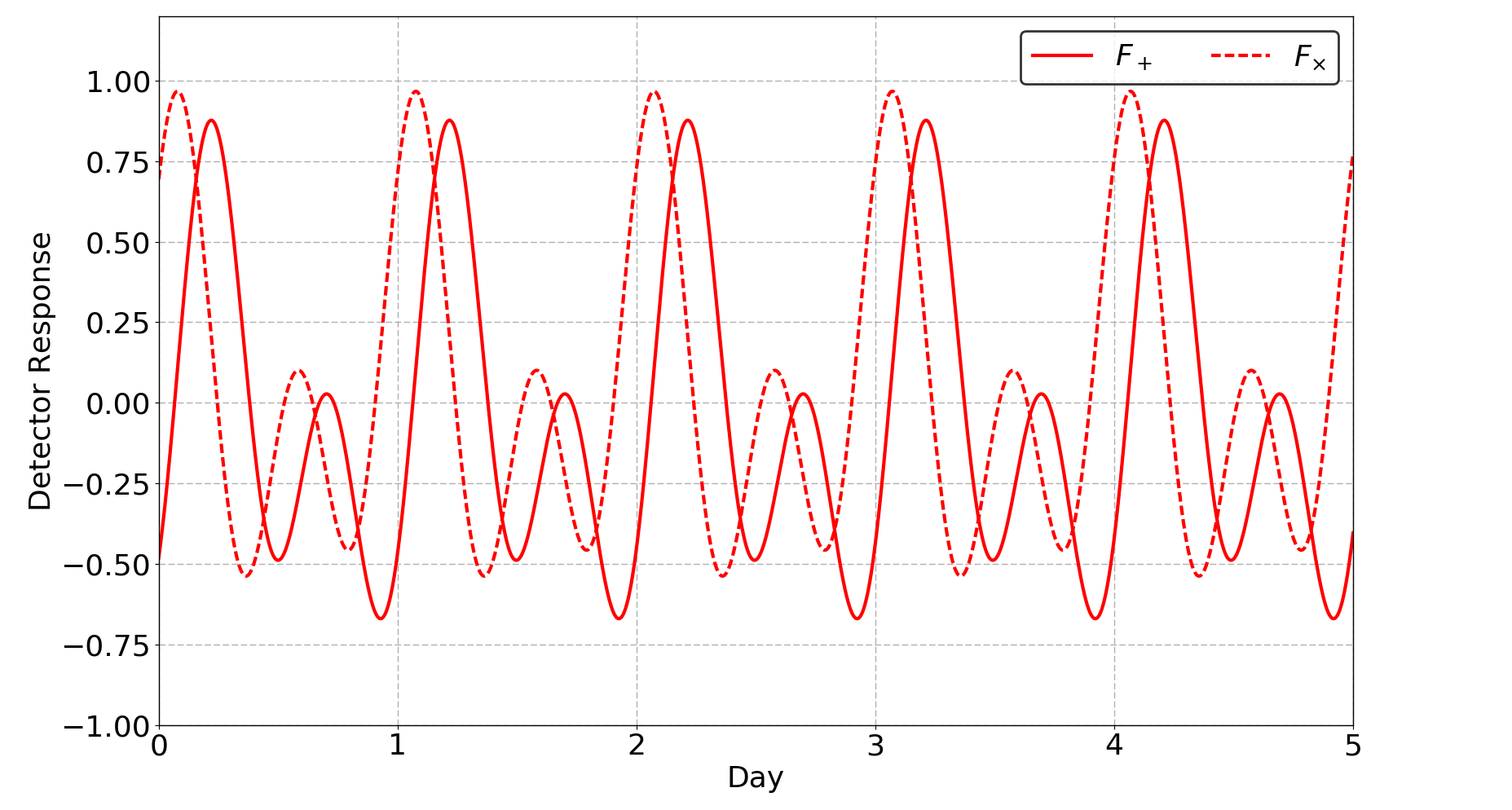}
        }%
    \end{center}
    \caption{
        The time-dependency of \ac{ET}'s and \ac{CE}'s detector response
to \acp{GW} with a polarization angle equal to $\pi/8$ coming from two example locations 
in the sky over the course of $5$ days.
Panel a and b are for the \ac{ET} and panel c and d for \ac{CE}.
Panel a and c show the detector response to a source located at $(\alpha$, $\delta)$ = 
$(0^{\circ}, 45^{\circ})$ and panel b and d show that to a source at $(\alpha$, $\delta)$ = 
$(30^{\circ}, 60^{\circ})$, where $\alpha$ and $\delta$ are right
ascension and declination of the source.
In the legend in panel a and b, the superscript $k = (1, 2, 3)$ indicates the $k^{\textrm{th}}$~
interferometer of the \ac{ET}.\label{fig:cycles}}%
\end{figure*}

%
Such a long duration allows the detector to observe the signal along the detector's 
trajectory on earth as the earth rotates, and therefore
makes the detector's response explicitly time-dependent. To illustrate this time-dependence, in Figure \ref{fig:cycles} two source sky locations are selected and the change over $5$ days of the \ac{ET} and \ac{CE} detector response to sources at those locations is shown.

%

\section{Methodology}\label{sec:method}
%
%
Using the Fisher matrix, we aim to provide a lower
bound on source sky position error for \ac{GW} sources and examine the feasibility of early warning.
It is often reported that the Fisher matrix approach produces estimates that are more optimistic than those methods that completely explore the likelihood such as \ac{MCMC}.  However, this is often due to the misuse of the Fisher matrix in situations where the \ac{SNR} is too low.  In the moderate to high \ac{SNR} regime where the Cramer–Rao lower bound is valid \cite{vallisneri2008use, zanolin2010application, cho2014application},
estimates from the Fisher matrix are a good indicator of the
expected uncertainty on parameters.  Furthermore, complete
\ac{MCMC} simulations are usually too computationally expensive
to carry out, while the Fisher matrix is low to moderate in computational cost. 

%
For an incoming \ac{GW}, the strain observed by the
$I^{\text{th}}$ detector can be expressed as $h_{I}(\boldsymbol{\theta}, t)$ in
the time domain.  It is a linear
combination of the wave's two polarizations $h_+(\boldsymbol{\theta},
t),~h_{\times}(\boldsymbol{\theta}, t)$ and the detector response
$F_{I}^+(\theta, \phi, \psi, t),~F_{I}^{\times}(\theta, \phi, \psi, t)$ as 
\ref{eq:data}
\begin{equation}\label{eq:data} 
h_{I}(\boldsymbol{\theta}, t) = F^+_{I}(\theta, \phi, \psi,
t)h_+(\boldsymbol{\theta}, t) + F^{\times}_{I}(\theta, \phi, \psi,
t)h_{\times}(\boldsymbol{\theta}, t), 
\end{equation}
where the vector $\boldsymbol{\theta}$ represents the unknown
signal parameters: sky position, distance,
time of arrival at the center of the earth, binary masses, initial phase
of the wave when it arrives at the center of the earth, inclination angle and
polarization angle. The time at the
detector is denoted by $t$ which is equal to the arrival
time $t_0$ of the incoming wave at the center of the earth, plus the time $\tau$ required for the wave to travel from the center of the earth to the detector, given by
\begin{equation}\label{eq:tdetector}
\tau = \frac{\bold{n} \cdot \bold{r}}{c},
\end{equation}
where $\bold{n}$ is the \ac{GW} propagation direction and $\bold{r}$ is the location vector of the detector
relative to the center of the Earth. 
The Fourier transform of $h_{I}(\boldsymbol{\theta}, t)$ is then
defined as
\begin{equation}\label{eq:ft}
\tilde{h}_{I}(\boldsymbol{\theta}, f) = \int_{-\infty}^{\infty}
h_{I}(\boldsymbol{\theta}, t) e^{-2i\pi ft} dt,
\end{equation}
%
%
The mathematical definition of the Fisher matrix is given by
\begin{equation}\label{eq:def}
\Gamma_{ij} = \sum_{I=1}^{N} 2  \int_0^{\infty} \frac{\frac{\partial \tilde{h}_I^*(\boldsymbol{\theta}, f)}{\partial \theta_i}\frac{\partial \tilde{h}_I(\boldsymbol{\theta}, f)}{\partial \theta_j} 
+ \frac{\partial \tilde{h}_I^*(\boldsymbol{\theta}, f)}{\partial \theta_j}\frac{\partial \tilde{h}_I(\boldsymbol{\theta}, f)}{\partial \theta_i}}{S_I(f)}df.
\end{equation}
where  $\partial \tilde{h}(\boldsymbol{\theta},
f)/\partial \theta_i$ is the partial derivative of
$\tilde{h}(\boldsymbol{\theta}, f)$ with respect to the $i^{\text{th}}$ unknown
parameter $\theta_{i}$. The power spectrum density
of the $I^{\text{th}}$ detector is denoted by $S_I(f)$.
We also sum over the number of detectors, or
in the case of the \ac{ET}, over the number of individual interferometers.  The
optimal \ac{SNR}, $\rho$, of the incoming \ac{GW} can be expressed as
\begin{equation}\label{eq:SNR}
\rho^2 = 4  \int_0^{\infty} \frac{|\tilde{h}_I{(\boldsymbol{\theta},
f)}|^2}{S_I(f)} df.
\end{equation}

%
In this work, we construct the Fisher matrix for the following
unknown parameters: right ascension, $\alpha$; declination, $\delta$; arrival time, $t_0$, at the center of the earth; the log of the distance, $\log_{10}d$; polarization angle, $\psi$; the log of the total binary masses, 
$\log_{10}M $; the cosine of the inclination angle, $\cos\iota$; 
the symmetric mass ratio, $\eta = M_1 \times M_2 / M^2$, of the masses of the two bodies in the binary; the initial phase, $\phi_0$, of the wave when it arrives at the center of the earth. 

%
When computing the \ac{GW} localization error 
for a source at a particular sky
location,  we divide the entire wave into pieces,
each of which is $100$ seconds long -- with the final piece
$\leq100$ seconds depending on the specific in-band duration of the
signal. The total number of pieces $N_\textrm{p}$ is then equal to
$\tau_\textrm{c}/100$, rounded towards positive infinity.
For each piece of the wave, we employ the formalism described above to compute
the Fisher matrix $\Gamma_{ij}^k$ and the optimal \ac{SNR} $\rho^{k}$. The
superscript $k$ indicates the $k^{\text{th}}$ piece of the wave. 
The final Fisher matrix $\Gamma_{ij}^{\textrm{f}}$ is then 
\begin{equation}\label{eq:sumFM}
\Gamma_{ij}^{\textrm{f}} = \sum_{k = 1} ^{N_p} \Gamma_{ij}^k,
\end{equation}
where we sum over the Fisher matrix contributions from each piece of the waveform, and the superscript $\textrm{f}$ indicates the resultant Fisher matrix.
The matrix inverse of the Fisher matrix then gives the covariance matrix of the unknown parameters as 
\begin{equation}\label{eq:invFM}
\text{cov}_{ij} = \Gamma_{ij}^{-1},
\end{equation}
from which the localization error is extracted using 
\begin{equation}\label{eq:eigen}
\Delta \Omega = 2 \pi \sqrt{\lambda_{\alpha} \lambda_{\delta}}\cos\delta,
\end{equation}
where $\Delta\Omega$ is the localization uncertainty, $\lambda_{\alpha}$ and $\lambda_{\delta}$ 
are the eigenvalues of the matrix $\text{cov}_{ij}$ corresponding to the $\alpha$ and $\delta$ of the source respectively.
The following expression can be used to convert $\Delta\Omega$ to any desired confidence level,
\begin{equation}\label{eq:eigen}
\Delta \Omega_{p} = -2\log(1-p)\Delta\Omega,
\end{equation}
where $p$ is a value between $0$ and $1$ indicating the confidence level.
Similarly, the accumulated \ac{SNR} is given by
\begin{equation}\label{eq:sumSNR}
(\rho^f)^2 = {\sum_{k=1}^{N_p} (\rho^{k})^2}.
\end{equation}

\section{Results of Simulation and Discussion}\label{sec:simre}

\subsection{Localization}\label{subsec:localisation}
\begin{figure*}
     \begin{center}
        \subfigure[]{%
            \label{fig:ET_1}
            \includegraphics[width=0.5\textwidth]{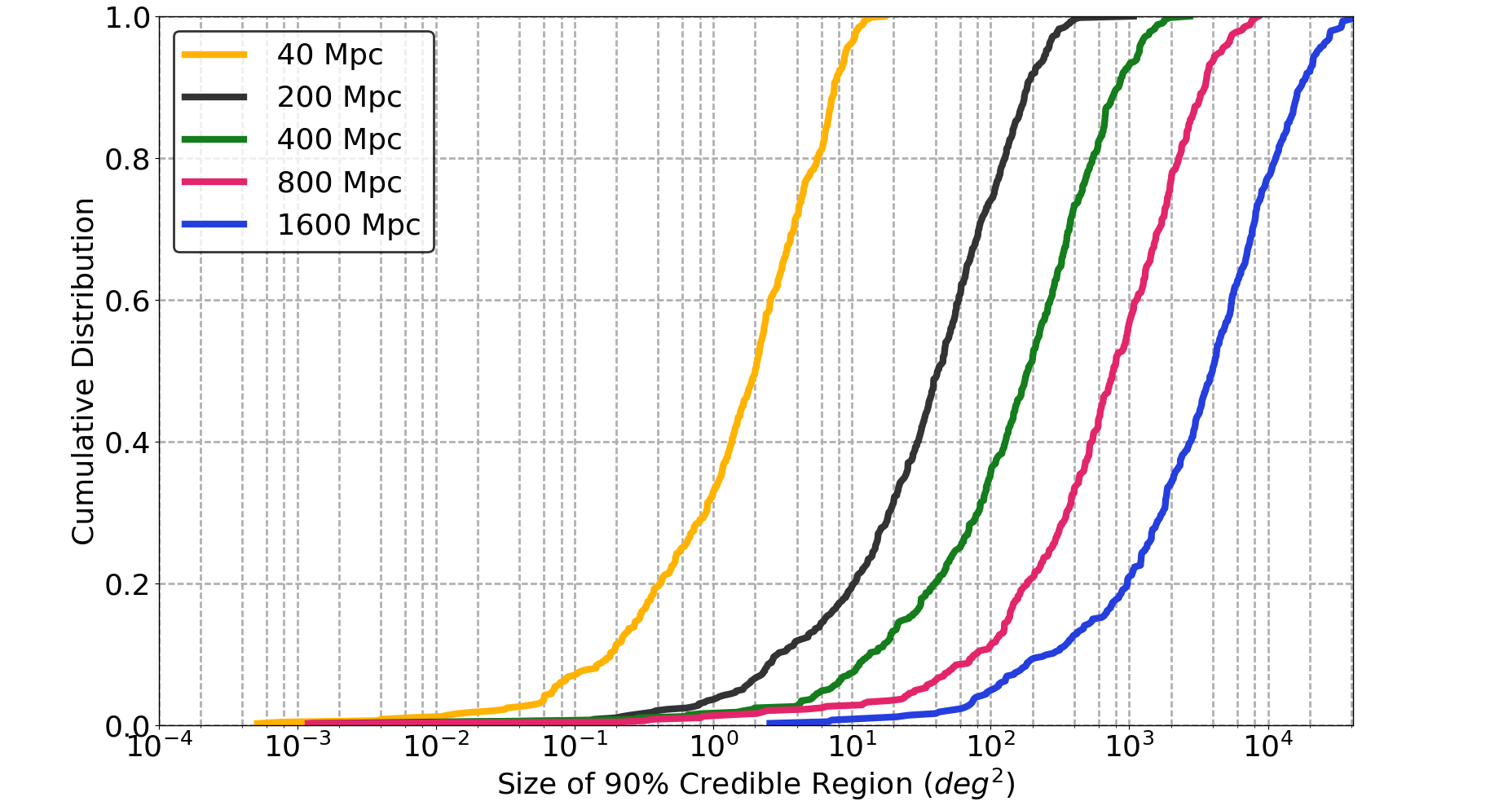}
        }%
        \subfigure[]{%
            \label{fig:ET_2}
            \includegraphics[width=0.5\textwidth]{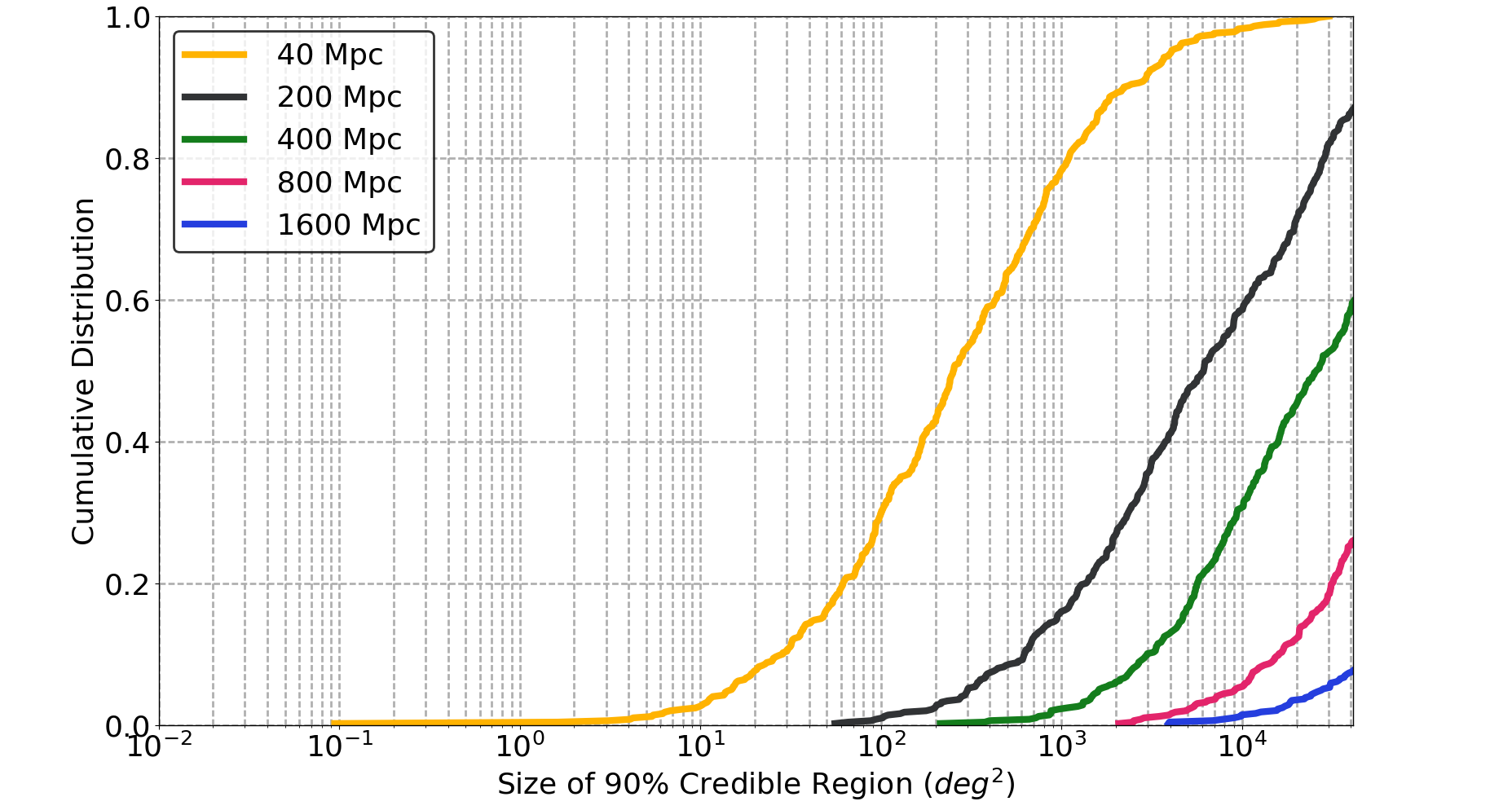}
        }%
	\\
	\subfigure[]{%
            \label{fig:CE_1}
            \includegraphics[width=0.5\textwidth]{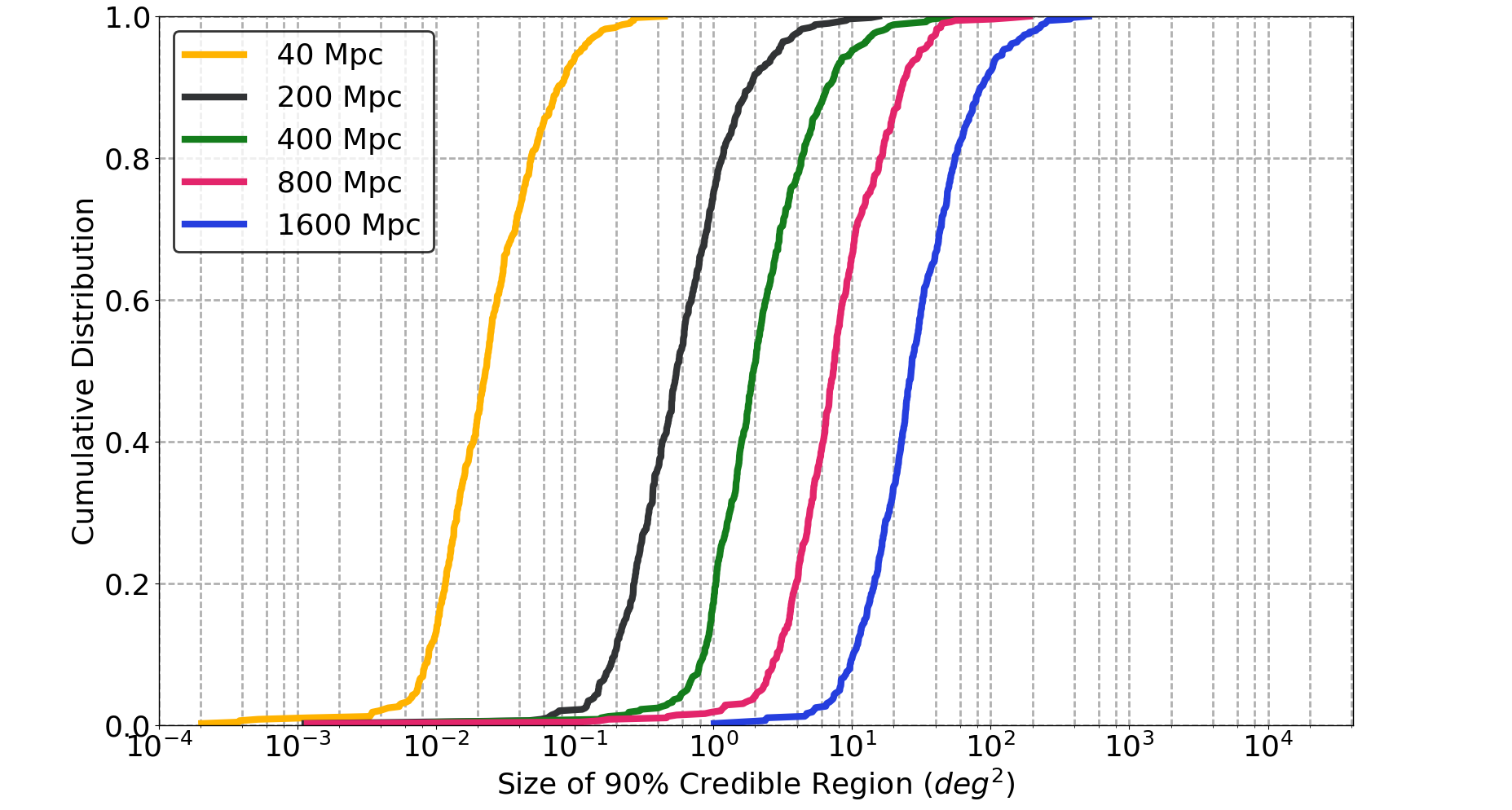}
        }%
    \end{center}
    \caption{\label{fig:BNS40} The cumulative distribution of the size of $90\%$
credible regions for sources at fixed distances. The
x-axes show the size of the $90\%$ credible region and the upper
limit of the x-axes corresponds to the size of the whole sky.
The yellow, black,  green,  red, and  blue lines represent
\ac{BNS} sources at $40$, $200$, $400$, $800$, and $1600$Mpc respectively. 
Panel a and b show the results for the \ac{ET} and \ac{CE} respectively. Panel c shows the results for the \ac{ET} 
and \ac{CE} as a network.}%
\end{figure*}

%
%
To test the localization capabilities of third generation detectors, we have
simulated \ac{GW} signals from $1.4M_{\odot}$-$1.4M_{\odot}$ \ac{BNS} sources at
distances of 40, 200, 400, 800, and 1600Mpc. The masses are defined in the local frame, i.e. $M_{\text{Local}}$,
which is related to the observed masses $M_{\text{Obs}}$ by
\begin{equation}\label{eq:masses}
M_{\text{Obs}} = M_{\text{Local}}(1 + z),
\end{equation}
We use $z$ to denote redshift.
All the masses defined earlier refer to the observed masses $M_{\text{Obs}}$.
The inclination angle $\iota$, polarization angle
$\psi$ and the sky position $(\alpha, \delta)$ are randomized. For each
specific distance, we have simulated $500$ \ac{BNS} signals. To determine
whether a source is detectable, we have employed different \ac{SNR} cuts for
each network configuration. For networks with more
than one interferometer such as the \ac{ET} or the
\ac{ET} and \ac{CE}, we have applied an \ac{SNR} requirement similar to that in
\cite{fairhurst2011source}. A detection is achieved if the network \ac{SNR} is
larger than or equal to 12 and the \acp{SNR}
in at least two interferometers are no less than $5.5$. For a network
with only one interferometer, namely, \ac{CE}, we
require that the accumulated \ac{SNR} is no less than
12 to claim a detection \footnote{We require a single detector to achieve 
\ac{SNR} $\geq 12$ for a detection for consistency. The results shown here for a single detector may therefore be more pessimistic than the reality.}.
The results of the simulations are presented as cumulative distributions in
Figures \ref{fig:BNS40}.

%
For  \acp{BNS} at $40$Mpc using only the \ac{ET}, $50\%$ of the
detectable sources can be localized with $90\%$ confidence to within
$2.0\text{deg}^2$, and $90\%$ of the detectable sources to within
$7.5\text{deg}^2$. For the best localized $10\%$ of sources, the
$90\%$ credible region is within
$0.2\text{deg}^2$ and these correspond to the
best located and orientated sources. For \acp{BNS} at $200$Mpc, $50\%$ and
$90\%$ of the detectable sources can be localized with $90\%$ confidence to within
$42\text{deg}^2$ and $183\text{deg}^2$ respectively.
Assuming \ac{EM} follow-up observations are achievable for sources that are
localized to within $100\text{deg}^2$, this indicates $100\% \, (74\%)$ of the
detectable sources at $40$Mpc ($200$Mpc), suggesting many opportunities for joint
\ac{EM} observations provided by the \ac{ET} for \acp{BNS} within $200$Mpc. 

%
For sources located at $400$Mpc, the upper limit of
the size of $90\%$ credible region increases to
$187\text{deg}^2\,(812\text{deg}^2)$ for the best localized $50\%$ ($90\%)$
of the detectable sources. 
This still leaves $36\%$ of the detectable sources localized
to within $100\text{deg}^2$ with $90\%$ confidence.  For sources located at
larger distances, i.e. $800$Mpc and $1600$Mpc, 
the upper limit of the size of $90\%$ credible regions for the best localized $50\%\,(90\%)$ of the detectable
sources increases substantially to $764\text{deg}^2$ ($3485\text{deg}^2$) and $3994\text{deg}^2$ ($1.7\times10^4\text{deg}^2$)
respectively. Moreover, only $11\%$ and $5\%$ of the
detectable sources can be localized to within $100\text{deg}^2$. This is
because the amplitude of the signals from sources at greater distances will be
weaker. Also, the observed $M_{\text{Obs}}$ as defined in Eq.\ref{eq:masses} will be
larger, meaning that the in-band duration will be shorter.
This suggests that localization of a \ac{BNS} at such distances by the \ac{ET} alone will still be poor and \ac{EM} follow-up observations will remain a challenge if the \ac{ET} is the only operating detector. 

Since the sensitivity of \ac{CE} at low
frequencies is limited, the in-band durations of the signals are shorter than that in the
\ac{ET}. As shall be seen later, the time-dependent modulation of detector response is
the main factor contributing to improved localization.
Consequently, the localization of a \ac{BNS} by \ac{CE} alone is worse.  For example,
$50\%\,(90\%)$ of the detectable \acp{BNS} at $40$Mpc can be localized to only within
$252\text{deg}^2\,(2212\text{deg}^2)$, a factor of $\sim126$ and $\sim 295$ larger than using only the \ac{ET}. 
Only $30\%$ of the detectable sources can be localized to within $100\text{deg}^2$ with $90\%$
confidence.
For sources at distances $\geq 400$Mpc, the upper limits of localization error
for the best localized $50\%$ and $90\%$ are larger than the whole sky.  This
means that for some sources, despite accumulating
enough \ac{SNR} to claim a detection, no localization information is available. 

%
Combining the \ac{ET} and \ac{CE} together as a network greatly improves 
the localization since it vastly increases the geographical baseline of the
network.  This greatly improves triangulation between the detectors in the network and will take advantage of the high frequency,
high \ac{SNR} components of the waveform, i.e., the final seconds. This will complement the localization information gained from the long duration and changing antenna patterns. All sources within $200$Mpc are
localized to within $30\text{deg}^2$ with $90\%$ confidence.  Importantly, at $40$Mpc and $200$Mpc, the $90\%$ credible region upper limit for the best localized $90\%$ of the detectable sources are only $\mathcal{O}(10^{-2})\text{deg}^2$ and $\mathcal{O}(1)\text{deg}^2$ respectively. 
For the detectable sources at $1600$Mpc, there are
still $92\%$ localized to within $100\text{deg}^2$ with $90\%$ confidence.
This shows a great promise for \ac{BNS}
multimessenger astronomy even at relatively large cosmological distances.


\begin{figure}
\includegraphics[width=0.5\textwidth]{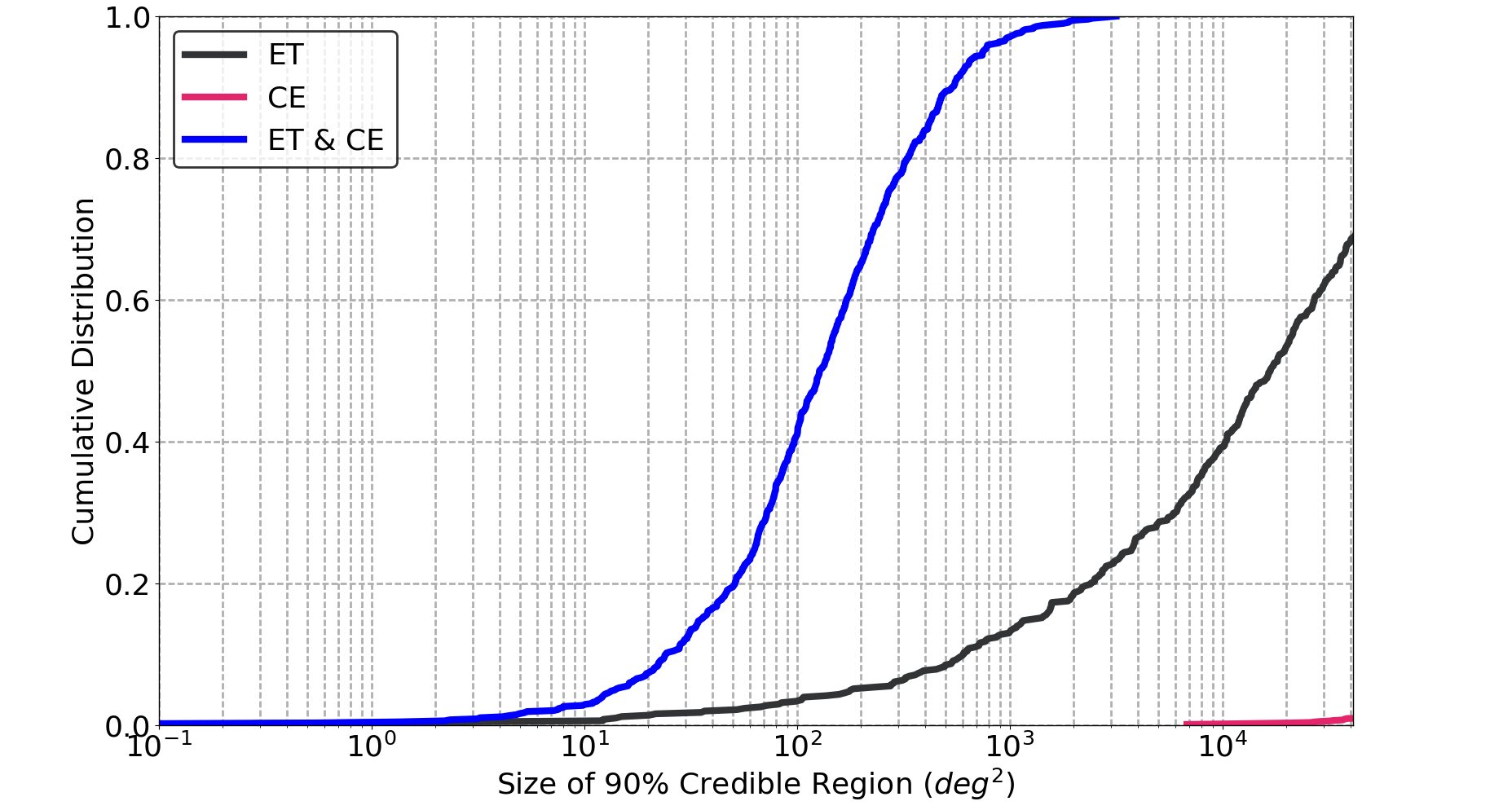}
\caption{The cumulative distribution of the size of $90\%$
credible regions in the sky, for detectable BNS sources
uniformly distributed in comoving volume, observed by the
\ac{ET} and \ac{CE} both individually and as a network. The
upper limit of the x-axis corresponds to the size of the entire
sky.\label{fig:ETCDF}} 
\end{figure}

%
To present a more general picture, we have also simulated the localization of a population of \ac{BNS} that are distributed uniformly in the comoving
volume. The results are presented in Figure~\ref{fig:ETCDF} as
cumulative distributions. Using the \ac{ET} alone, 
the farthest detectable source is at $z = 1.7$. Of the
detectable sources, $50\%$ can be localized to within $\sim1.7 \times
10^4\text{deg}^2$ with $90\%$ confidence. 
The cumulative distribution for the \ac{ET} reaches $68\%$
when the value at the x-axis is the size of the entire sky -- i.e.
this indicates that for up to $32\%$ of the detectable
sources, essentially no localization information is available. 
For \ac{CE}, the situation is worse. The farthest detectable source is located at $z = 4.9$, 
but only $\sim2\%$ of the detectable sources will have any localization information available. 
Again, a network with the \ac{ET} and \ac{CE} can bring a
huge improvement to the localization performance. 
For example, compared to using the \ac{ET} only, the
upper limit of the $90\%$ credible region for the best localized $90\%$ of the detectable sources
has been reduced by a factor $>100$ to $123\text{deg}^2$. 
The fraction of detectable sources that can be
localized to within $100\text{deg}^2$ with $90\%$ confidence has increased by
more than $10$ times to $43\%$. Interestingly, the farthest detectable source with a network of the \ac{ET} 
and \ac{CE} is located at $z = 2.2$. This is because for a network of more than one interferometer, we require an \ac{SNR} $\geq 5.5$ in at least two of the interferometers, besides also requiring a network \ac{SNR} $\geq 12$.  For sources located at $z > 2.2$, only the \ac{CE} is able to accumulate enough \ac{SNR} -- leading to a failure to meet the detection criterion.

A summary of the results is given in Table~\ref{table:summary}.  
Given the success of the \ac{EM} follow-up observations of GW170817, where the
localization error at $90\%$ confidence is $28\text{deg}^2$
\cite{PhysRevLett.119.161101}, also presented in the 
table is a column showing the percentage of detectable
sources that can be localized to within $30\text{deg}^2$ with $90\%$ confidence.
\begin{table}
\begin{threeparttable}
\caption{Statistical Summary of Results}
\label{table:summary}
\begin{tabular}{ccccccc}
\\
\toprule
{\multirow{2}{*}{Network}}              & d& {\multirow{2}{*}{$n$}} & $50\% $ & $90\%$ & $\leq 100$ & $\leq 30$ \\
                                          & (Mpc) & & $(\text{deg}^2) $ & $(\text{deg}^2)$& $(\text{deg}^2)$ &$(\text{deg}^2)$ \\ \hline 
{\multirow{6}{*}{ET}}  & 40 &  {\multirow{5}{*}{500}}  & $2$  & $8$ &$100\%$ & $100\%$ \\
                        & 200 &  & $42$& $183$ &  $74\%$&$40\%$\\
                         & 400 & & $187$& $ 837$ & $36\%$ &$16\%$\\
                          & 800 &   &$764$ &$3485$ & $11\%$ &$5\%$\\
                           & 1600 &   &$3994$ & $1.7 \times 10^4$& $5\%$ &$2\%$\\
 &          Uniform $^{1}$  &  3000 & $1.7 \times 10^4$ & $> \textrm{Sky}$& $3\%$&$2\%$ \\ \hline
 {\multirow{6}{*}{CE}}  & 40 &  {\multirow{5}{*}{500}}  & $252$  & $2212$ &$30\%$ & $10\%$ \\ 
                        & 200 &  & $6118$& $> \textrm{Sky}$ &  $1\%$& $0\%$\\
                         & 400 & & $2.6 \times 10^4$& $> \textrm{Sky}$ & $0\%$ &$0\%$\\
                          & 800 &   &$> \textrm{Sky}$ &$> \textrm{Sky}$ & $0\%$ &$0\%$\\
                           & 1600 &   &$> \textrm{Sky}$ & $> \textrm{Sky}$& $0\%$ &$0\%$\\
 &          Uniform $^{1}$  &  5000 & $> \textrm{Sky}$ & $> \textrm{Sky}$& $0\%$&$0\%$ \\  \hline
 {\multirow{6}{*}{ET $\&$ CE}}  & 40 &  {\multirow{5}{*}{500}}  & $2\times10^{-2}$  & $8\times10^{-2}$ &$100\%$ &$100\%$ \\ 
                        & 200 &  & $5\times10^{-1}$& $1.8$ &  $100\%$&$100\%$\\
                         & 400 & & $2$& $ 7$ & $100\%$ &$99\%$\\
                          & 800 &   &$7$ &$23$ & $99\%$ &$94\%$\\
                           & 1600 &   &$27$ & $85$& $92\%$ &$55\%$\\
 &          Uniform $^{1}$  &  5000 & $128$ & $538$& $41\%$&$12\%$ \\ 
\hline 
\end{tabular}
$^{1}$\scriptsize{Uniformly distributed in the comoving volume.}
\\
\begin{tablenotes}
\setlength\labelsep{0pt}
\normalfont{
\item A brief statistical summary of our results for sky localization. In
the first row, we use $d$ to denote distance and $n$
the number of injections. The third and the fourth
columns indicate the upper limit of the size of $90\%$ credible regions for the best localized $50\%$ and $90\%$ of the
detectable sources. The fifth column shows the percentage of the detectable
sources that can be localized to within $100\text{deg}^2$ with $90\%$
confidence, and the last column the percentage within $30\text{deg}^2$ with
$90\%$ confidence.}
\end{tablenotes}
\end{threeparttable}
\end{table}

\subsection{Early Warning}
%
%
In the era of third generation detectors, due to the extended
in-band duration of detectable signals, it is possible that signals will accumulate \ac{SNR} such that the trigger may be considered significant before the merger occurs. In
this section we investigate the feasibility of issuing early warnings prior to
binary coalescence. We assume that if the \ac{SNR} for a \ac{GW} event
can be accumulated before merger, up to a
level that satisfies the detection requirement as defined in
Section~\ref{subsec:localisation}, the event will be deemed significant. As the purpose of releasing an early warning is to increase the chance of successful \ac{EM} follow-up observation, releasing an alert too
early may result in a localization error too large to carry out any meaningful
follow-ups. We therefore require two criteria to be met before an alert can be released. Firstly, the signal has to satisfy the \ac{SNR} requirement for detection and secondly, the $90\%$ credible region has to be no larger than $100\text{deg}^2$ at the moment the alert is sent. We will refer to these two requirements as early warning criteria in the remaining of this paper.  As early warning is mostly made possible due to the improvement in the sensitivity in the low frequency band, we focus our analysis on the \ac{ET}, and the \ac{ET} and \ac{CE} as a network. The \ac{BNS} systems are distributed at specific distances and uniformly in the comoving volume as discussed before.

\begin{figure*}
\centering     
\subfigure[]{\label{fig:a}\includegraphics[width=\columnwidth]{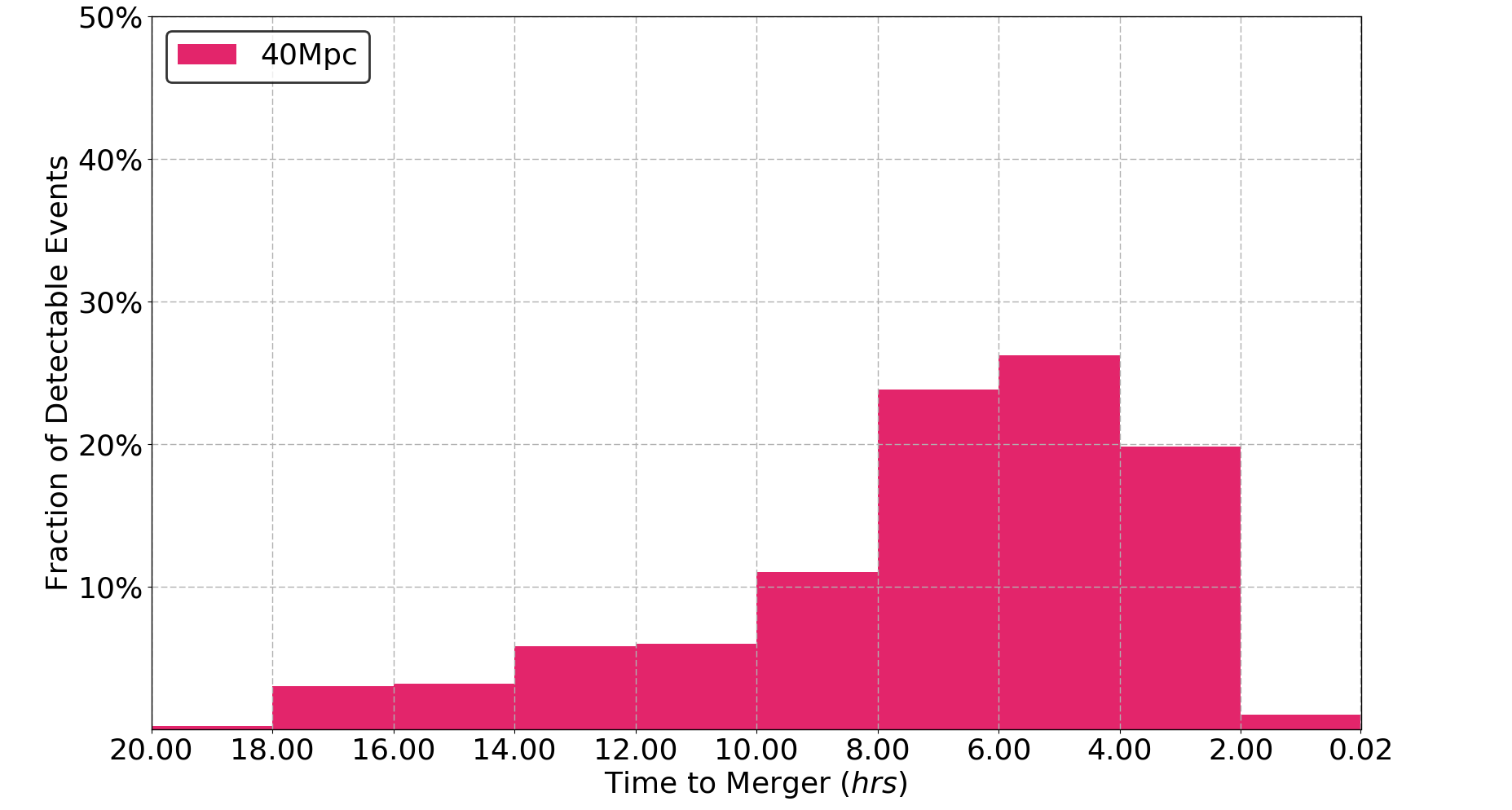}}
\subfigure[]{\label{fig:b}\includegraphics[width=\columnwidth]{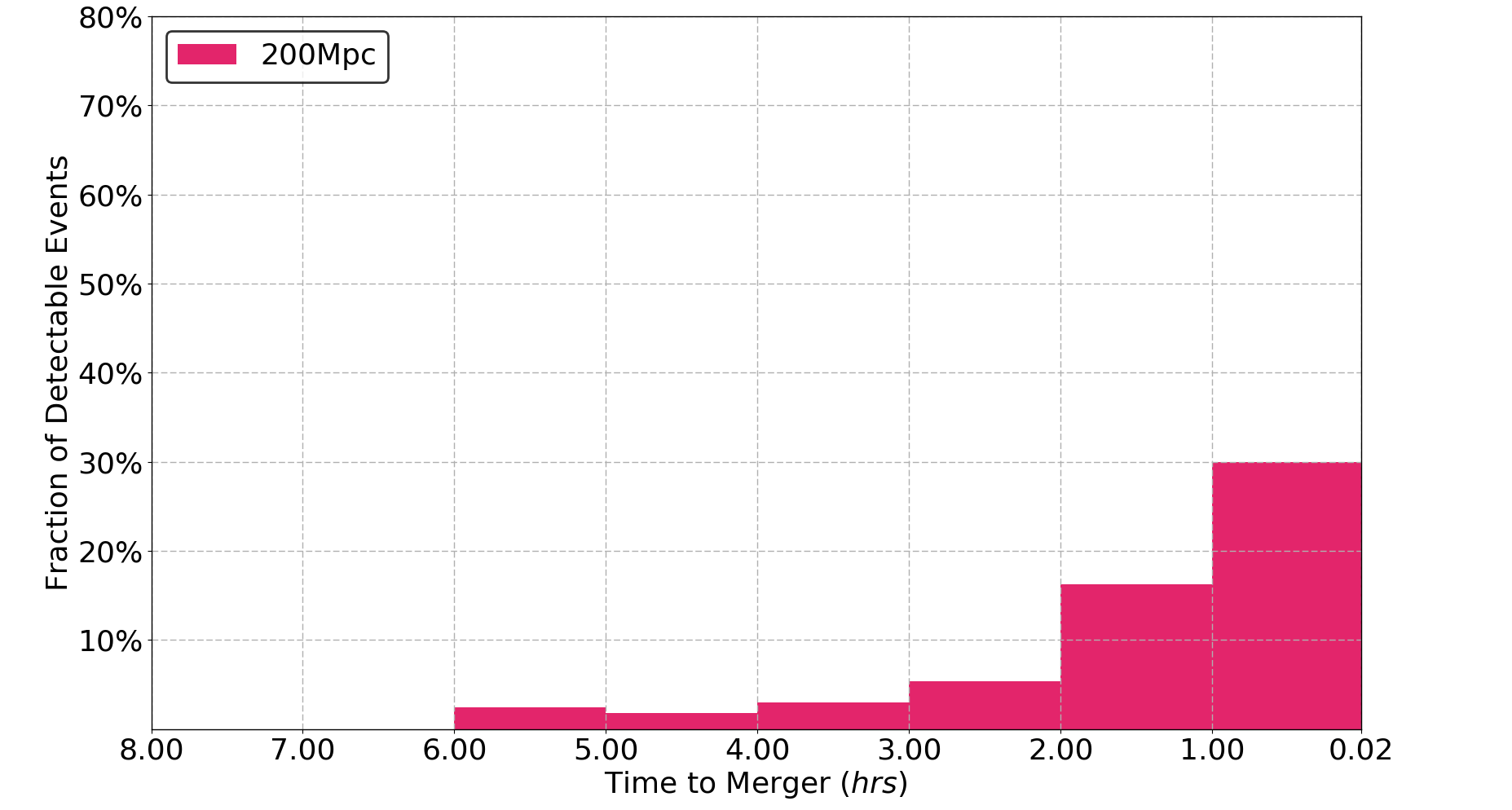}}\\
\subfigure[]{\label{fig:b}\includegraphics[width=\columnwidth]{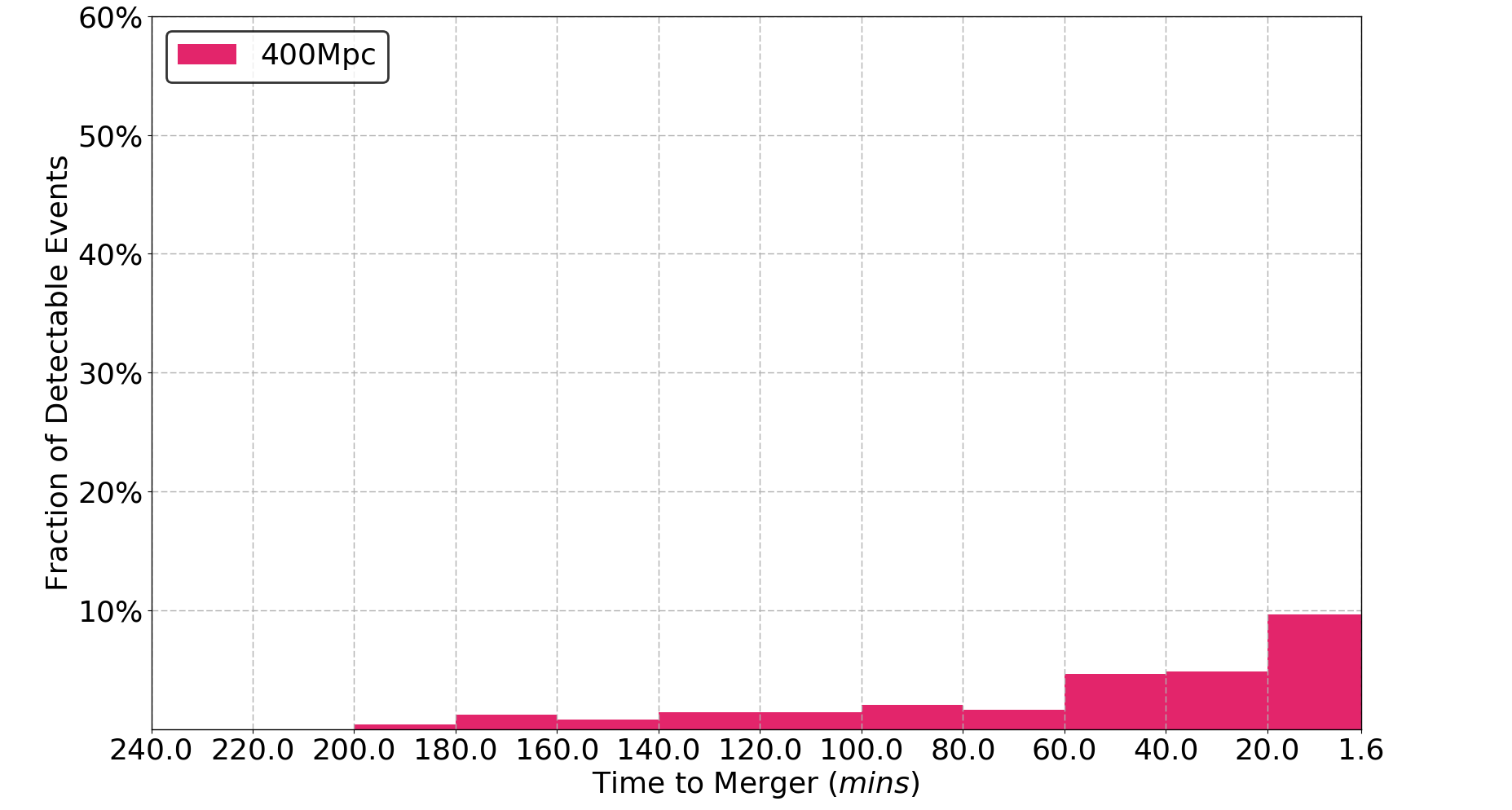}}
\subfigure[]{\label{fig:b}\includegraphics[width=\columnwidth]{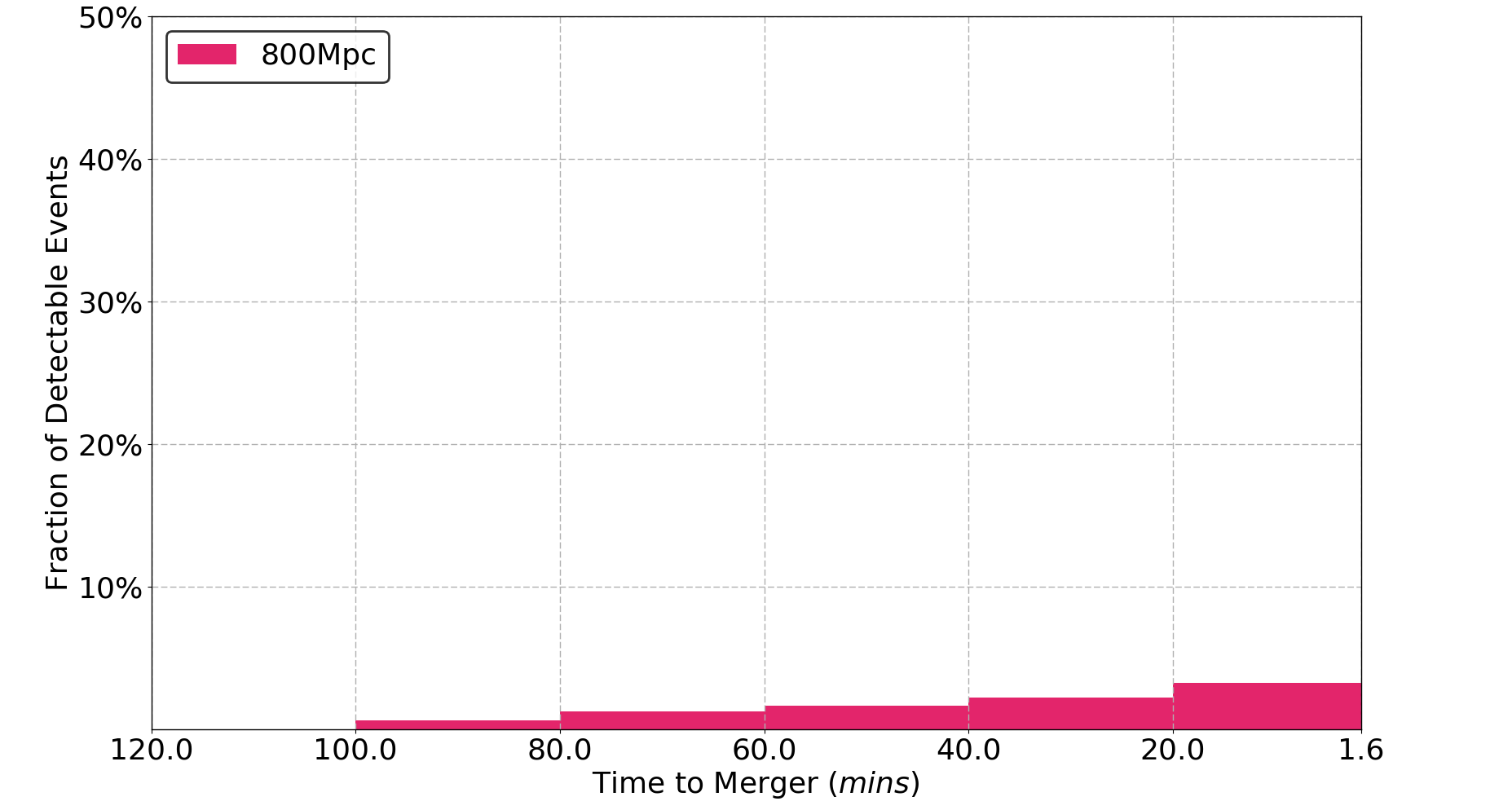}}\\
\subfigure[]{\label{fig:b}\includegraphics[width=\columnwidth]{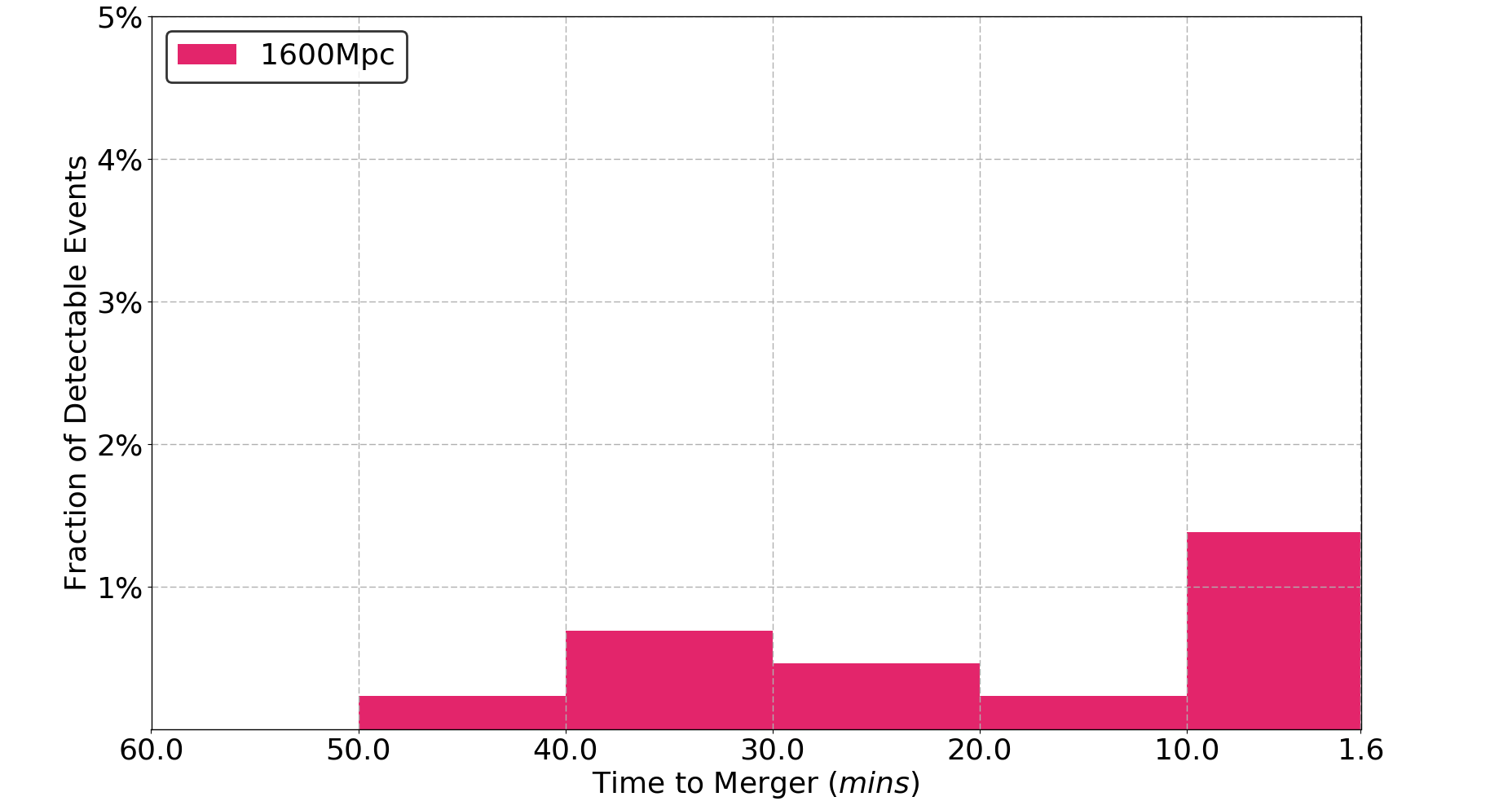}}
\caption{Histograms of the fraction of detectable
events that achieve the early warning criteria as a function of time to merger
for the ET detector. Panel a, b, c, d, and e are for events at $40$, $200$, $400$, $800$, and $1600$Mpc respectively. The x-axes indicate the time to merger when the signal meets the early warning
criteria. The y-axes indicate the fraction of detectable events that achieve
these early warning criteria. Note that at distances $\geq 400$Mpc,
since a large fraction of the times until merger will fall
within $1$ hour, for greater clarity the scale of the axes varies from panel to panel. Only those signals which achieve the early warning
criteria at least $100$ seconds prior to merger will be counted.\label{fig:ETEW}}
\end{figure*}

\begin{figure*}
\centering     
\subfigure[]{\label{fig:a}\includegraphics[width=\columnwidth]{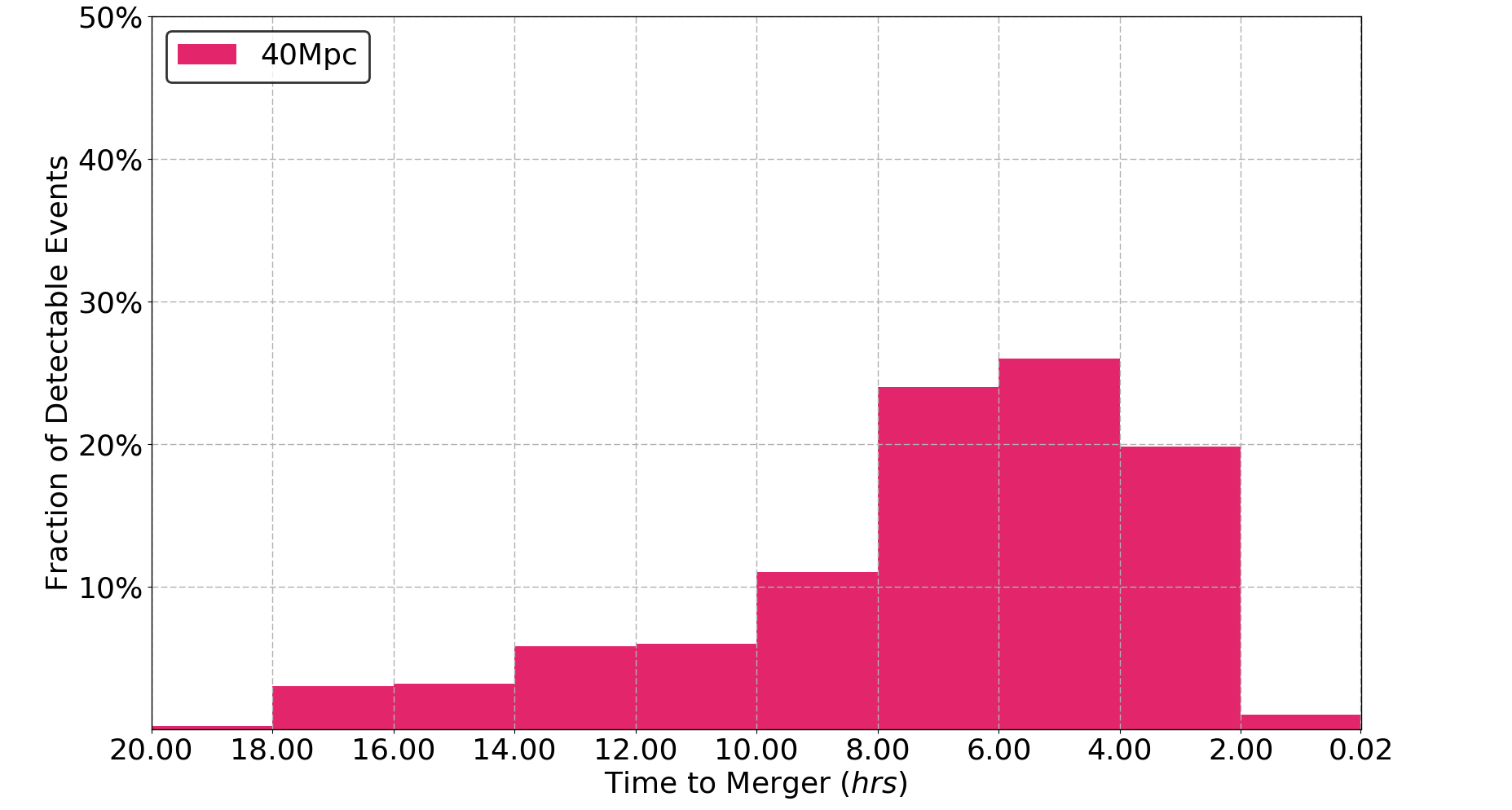}}
\subfigure[]{\label{fig:b}\includegraphics[width=\columnwidth]{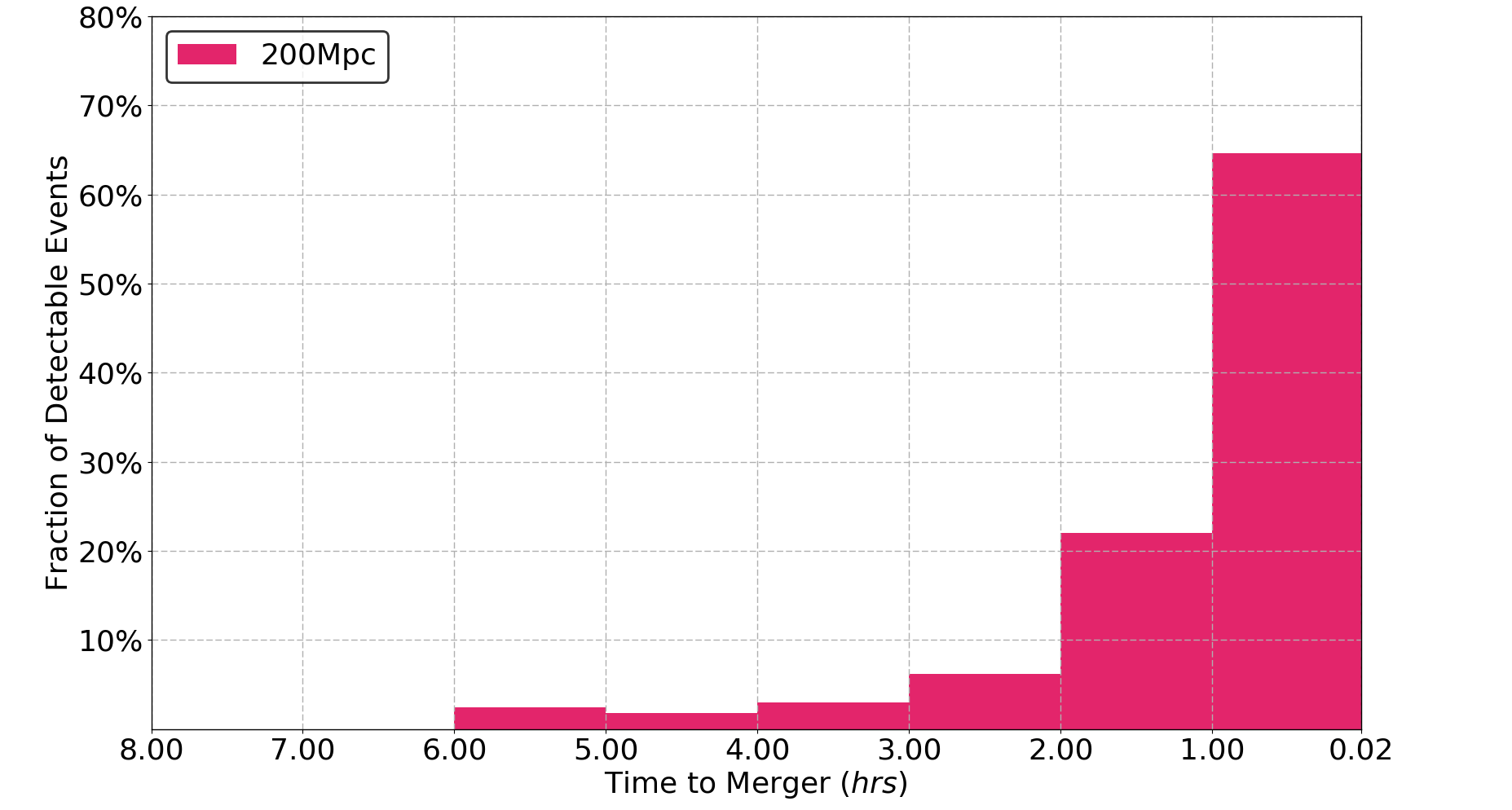}}\\
\subfigure[]{\label{fig:b}\includegraphics[width=\columnwidth]{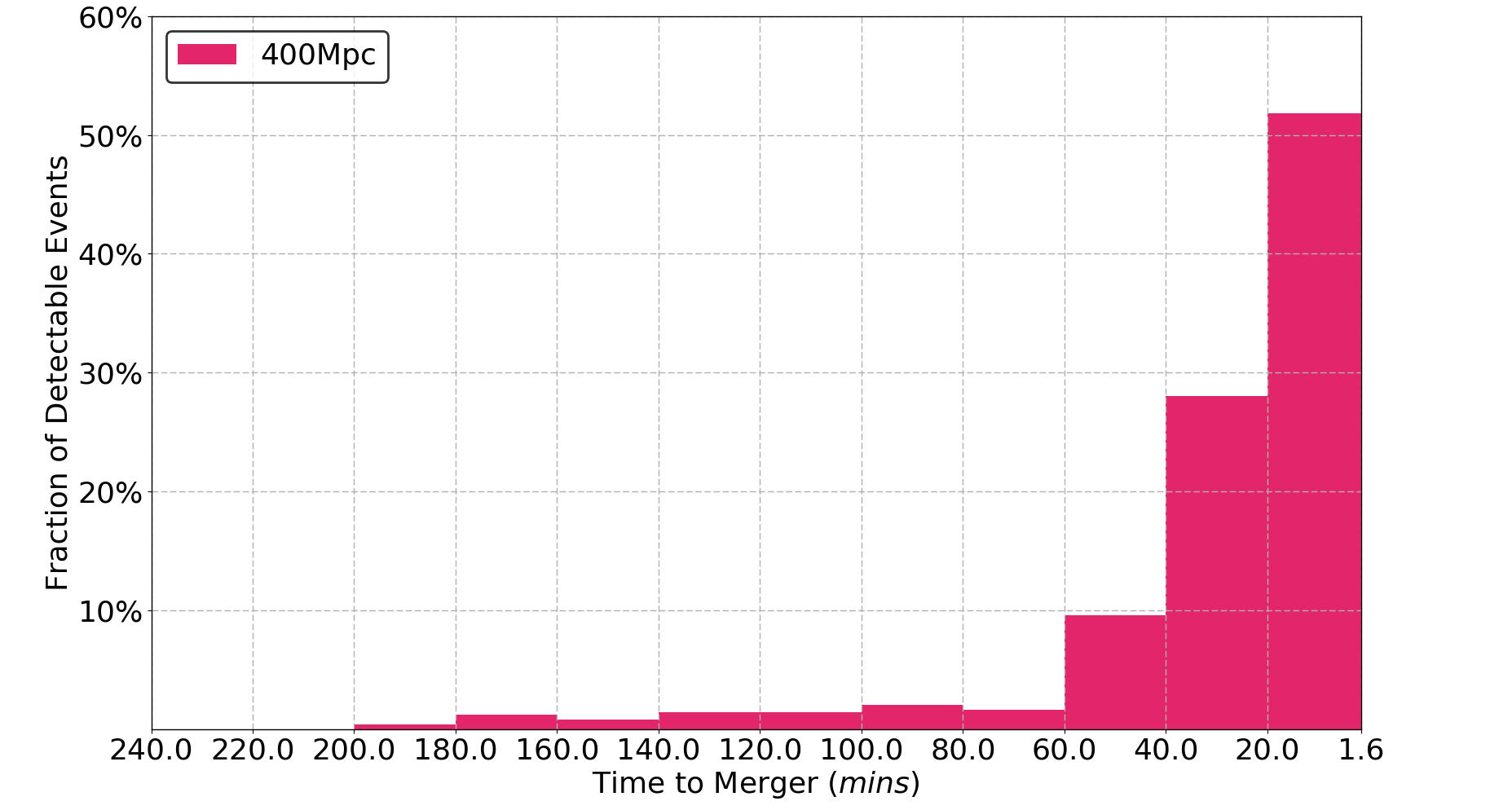}}
\subfigure[]{\label{fig:b}\includegraphics[width=\columnwidth]{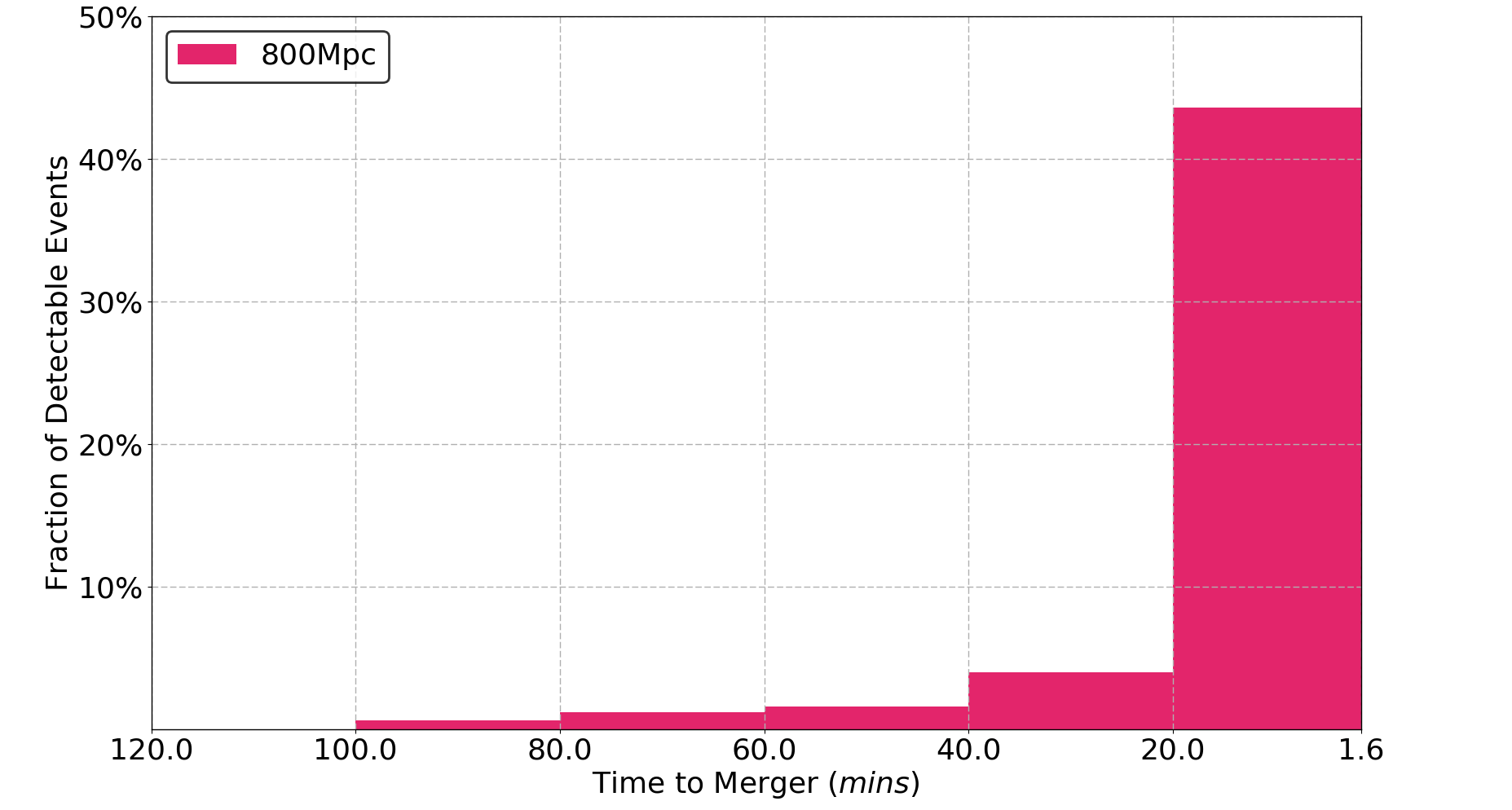}}\\
\subfigure[]{\label{fig:b}\includegraphics[width=\columnwidth]{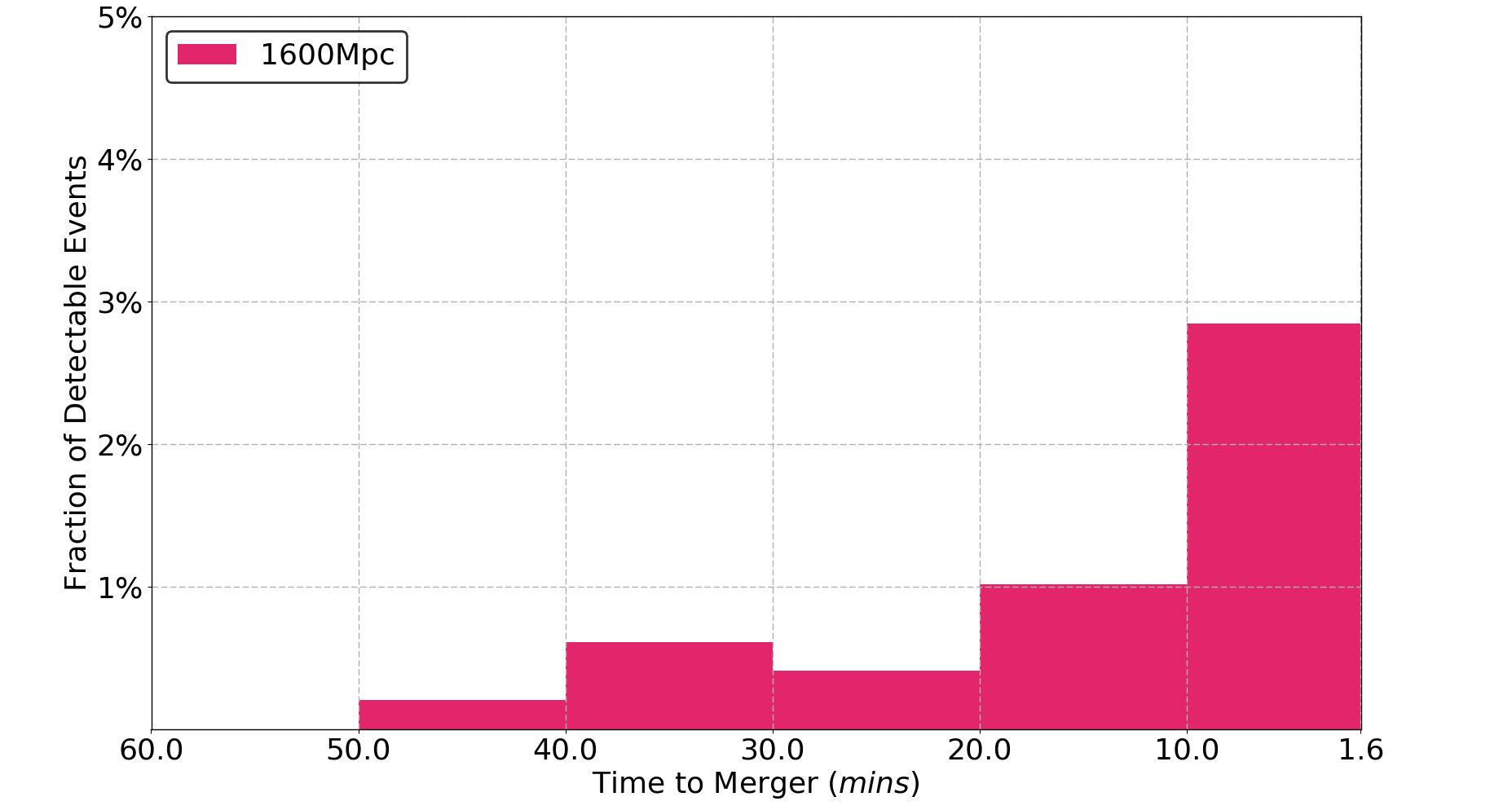}}
\caption{The same as Figure \ref{fig:ETEW} but using the \ac{ET}
and \ac{CE} as a network. For the same source distance, the scale of the panels is the same as is shown in Figure \ref{fig:ETEW}, to allow a
convenient comparison.
\label{fig:CEETEW}}
\end{figure*}

%
We present the results for the \ac{ET} in Figure~\ref{fig:ETEW} and the
\ac{ET} and \ac{CE} operating together as a network in Figure~\ref{fig:CEETEW}. These histograms show the distribution of the fraction of
detectable events as a function of the time before merger at which the events meet the early warning criteria.
Using the \ac{ET}, all the signals at $40$Mpc meet the early warning criteria
between $1$ and $20$ hours before merger, with the mode of the distribution at $\sim5$ hours. 
At $200$Mpc, $58\%$ of the detectable signals have accumulated enough \ac{SNR} for early warning between $1$
to $6$ hours prior to merger. This represents a
significant advantage that can be provided by the \ac{ET} in \ac{EM}
follow-up observations for sources within $200$Mpc.
As the distance increases, the fraction of detectable sources that meet the early warning criteria continues to drop.
Of the detectable sources at $400$Mpc,  only $\sim 27\%$ can meet the early warning criteria and the fraction further drops to $\sim 9\%$ and $\sim 3\%$ for sources at $800$Mpc and $1600$Mpc respectively.
Moreover, at $1600$Mpc, the times prior to merger when the signals meet the early warning criteria drop to $\leq50$ minutes.

As would be expected, an additional third generation detector
will improve the performance significantly and provide much improved early warning capability. 
In Figure~\ref{fig:CEETEW}, it can be seen that for distances $\geq200$Mpc and $\leq 1600$Mpc,
the distributions of early warning times have become
noticeably skewed to larger times compared to using only the
\ac{ET}. This suggests that a network of the \ac{ET} and \ac{CE} detectors
will provide better early warning capability for sources at relatively large distance. 
For example, the fractions of the detectable sources at $400$Mpc, $800$Mpc and $1600$Mpc
that can meet the early warning criteria are $98\%$, $51\%$ and $5\%$ respectively.
At $40$Mpc, since using only the \ac{ET} all the
sources will have already met the early warning criteria at a time when
the frequency of the signal is still relatively low, an additional detector of
\ac{CE} does not alter the distribution significantly.
At $1600$Mpc, the result may seem to suggest that a network of the \ac{ET} and \ac{CE} does not perform much better 
than using the \ac{ET} alone.
However, this is because a network of the \ac{ET} and \ac{CE} 
will be able to detect sources that are undetectable to the \ac{ET} alone.
These sources will not contribute much to the number of events that meet the early warning criteria but will contribute to the number of detectable events.

\begin{figure}
\centering     
\subfigure[]{\label{fig:a}\includegraphics[width=\columnwidth]{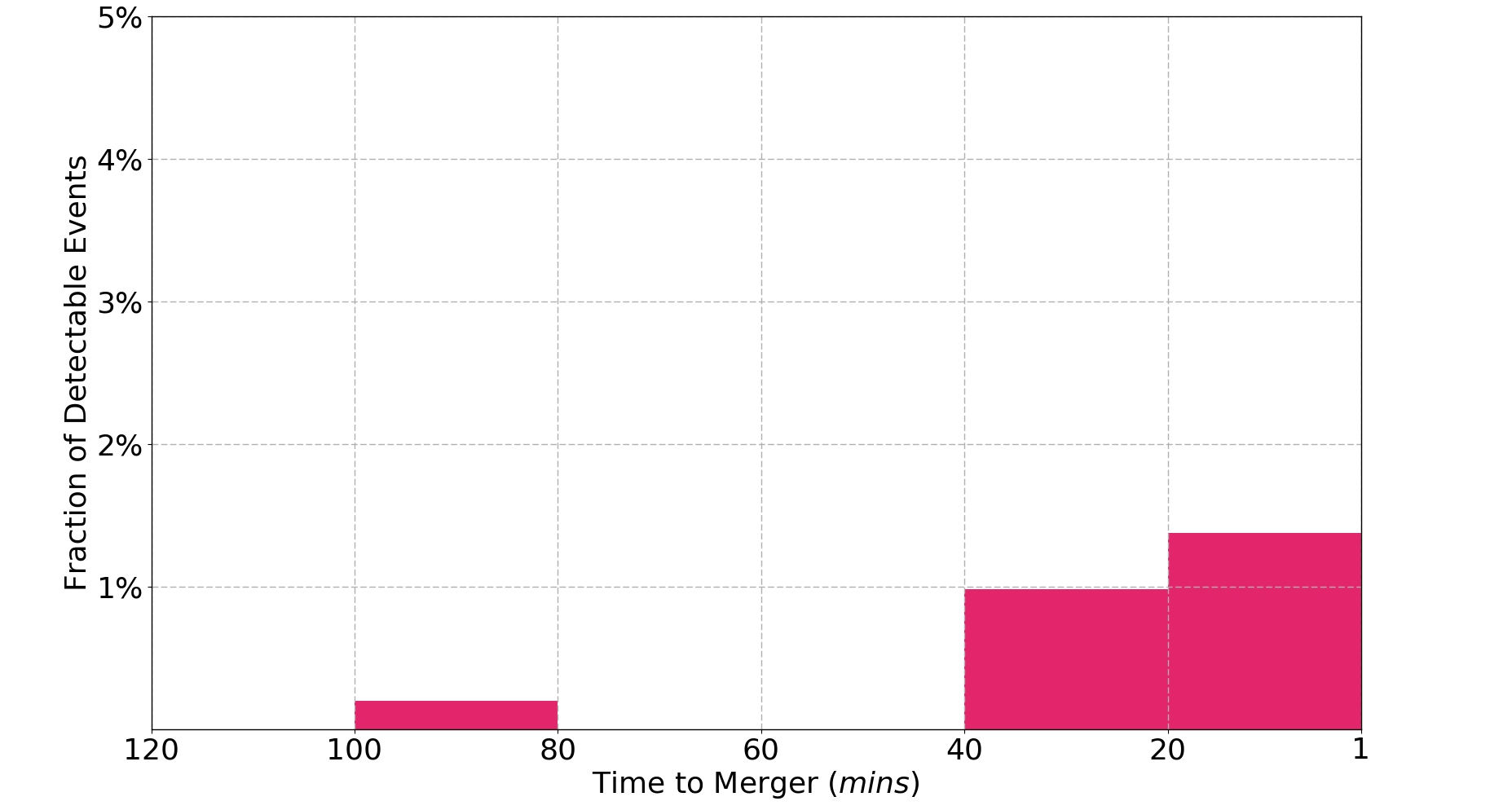}}
\subfigure[]{\label{fig:b}\includegraphics[width=\columnwidth]{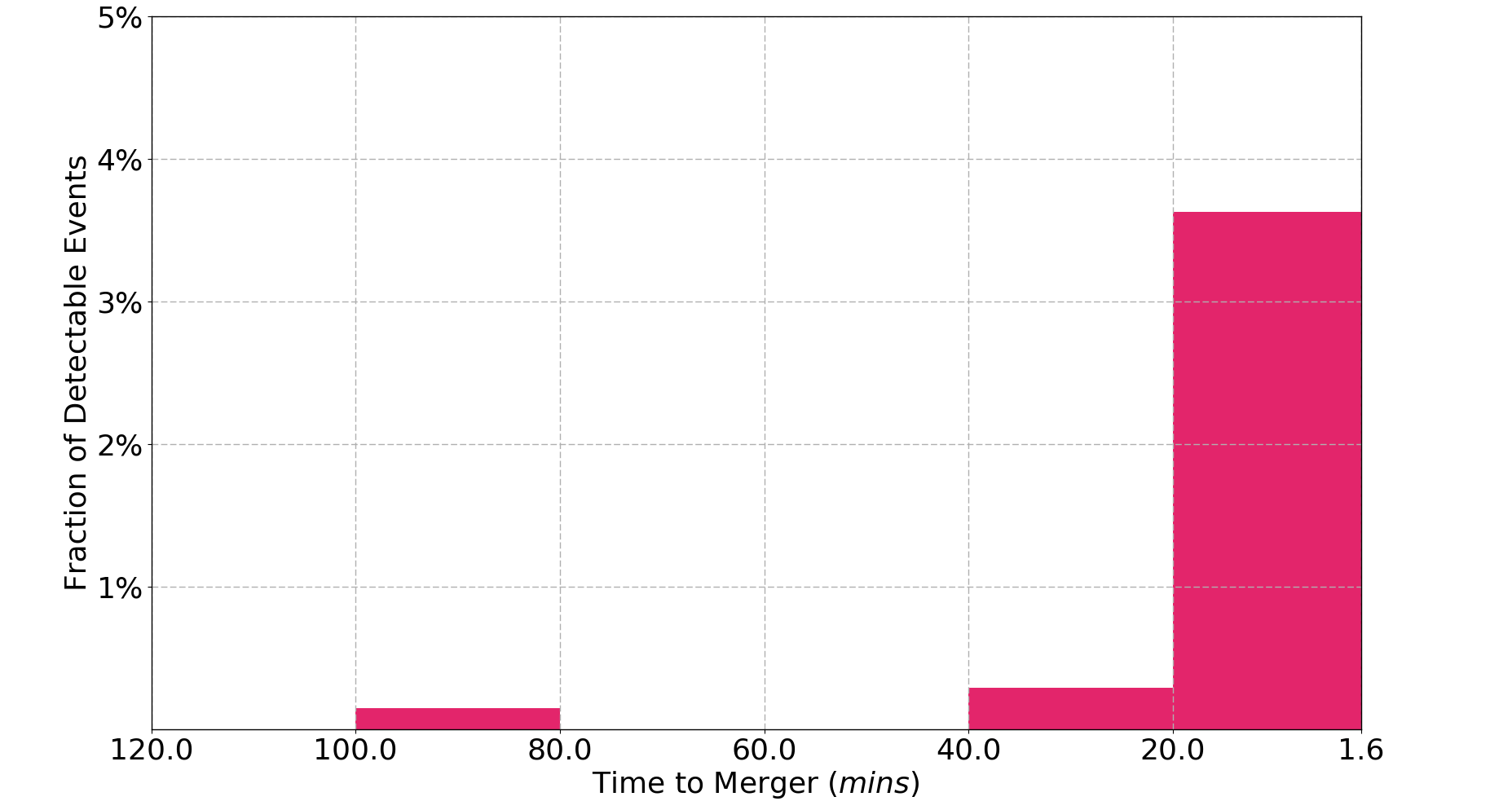}}
\caption{Histograms showing the fraction of detectable
events that meet the early warning criteria as a function of time to merger for a population of \ac{BNS} sources distributed uniformly in comoving volume.  Panel a shows the results for the \ac{ET} and
panel b for the \ac{ET} and \ac{CE} as a network.\label{fig:ETCEtEW}}
\end{figure}

Finally, to provide a more general picture, we present in Figure \ref{fig:ETCEtEW} the results for a population of \ac{BNS} distributed uniformly in comoving volume.
In line with the results shown in the previous sections, a network of the \ac{ET} and \ac{CE} will increase the number of events that meet the early warning criteria. With the \ac{ET} alone, $\sim 2\%$ of detectable sources can have their alerts released prior to merger. This ratio is $\sim 4\%$ after \ac{CE} joining the observation. However, the reason for the small increase in the fraction is because a network of the \ac{ET} and \ac{CE} will be able to detect sources that are undetectable to the \ac{ET} alone, and sources located at greater distances.
In Table \ref{table:summaryEW}, 
we present a summary of the results in terms of early warning.
\begin{table}
\begin{threeparttable}
\caption{Statistical Summary of Results For Early Warning}
\label{table:summaryEW}
\begin{tabular}{cccccccc}
\\
\toprule
{\multirow{2}{*}{Network}}              & d& {\multirow{2}{*}{$n$}} & $100 $ & $0.5 $ & $2$ & $5$ & $10$ \\
                                          & (Mpc) & & sec & hrs & hrs& hrs & hrs \\ \hline 
{\multirow{6}{*}{ET}}  & 40 &  {\multirow{5}{*}{500}}  & $100\%$ & $100\%$  & $99\%$ &$66\%$ & $18\%$ \\
                        & 200 &  & $58\%$&$39\%$& $13\%$ &  $2\%$&$0\%$\\
                         & 400 & & $28\%$& $16\%$& $ 4\%$ & $0\%$ &$0\%$\\
                          & 800 &   & $9\%$&$4\%$ &$0\%$ & $0\%$ &$0\%$\\
                           & 1600 &   & $3\%$&$1\%$ & $0\%$& $0\%$ &$0\%$\\ 
&          Uniform $^{1}$  &  3000 & $2\%$ & $1\%$& $0\%$&$0\%$&$0\%$ \\  \hline
                           
 {\multirow{6}{*}{ET $\&$ CE}}  & 40 &  {\multirow{5}{*}{500}}  &$100\%$&  $100\%$  & $99\%$ &$66\%$ &$18\%$ \\ 
                        & 200 &  & $100\%$& $74\%$& $13.4\%$ &  $2\%$&$0\%$\\
                         & 400 & & $98\%$& $27\%$& $ 4\%$ & $0\%$ &$0\%$\\
                          & 800 &   &$51\%$ &$4\%$ & $0\%$ &$0\%$ &$0\%$\\
                           & 1600 &   &$5\%$ &$1\%$ & $0\%$& $0\%$ &$0\%$\\
 &          Uniform $^{1}$  &  5000 & $4\%$ & $1\%$& $0\%$&$0\%$ & $0\%$ \\ 
\hline 
\end{tabular}
$^{1}$\scriptsize{Uniformly distributed in the comoving volume.}
\\
\begin{tablenotes}
\setlength\labelsep{0pt}
\normalfont{
\item A brief statistical summary of the results for early warning. In
the first row, we again use $d$ to denote distance and $n$
the number of injections. The third to the seventh columns 
indicate the fraction of detectable events that meet the early warning criteria within the corresponding times.}
\end{tablenotes}
\end{threeparttable}
\end{table}

As discussed, modulations of the Doppler effect and time-dependent detector responses are the two main consequences that will be seen 
in long in-band duration signals. 
Zhao and Wen, 2017 \cite{zhao2017localization} 
has tested thoroughly the difference in localizations with or without including the time-dependencies of these two effects, for networks of third generation detectors. However, it is still not clear which of these two factors has a more important role in terms of localizing \ac{BNS} mergers.
We here investigate the relative importance of these two factors.

\begin{figure}
\includegraphics[width=0.5\textwidth]{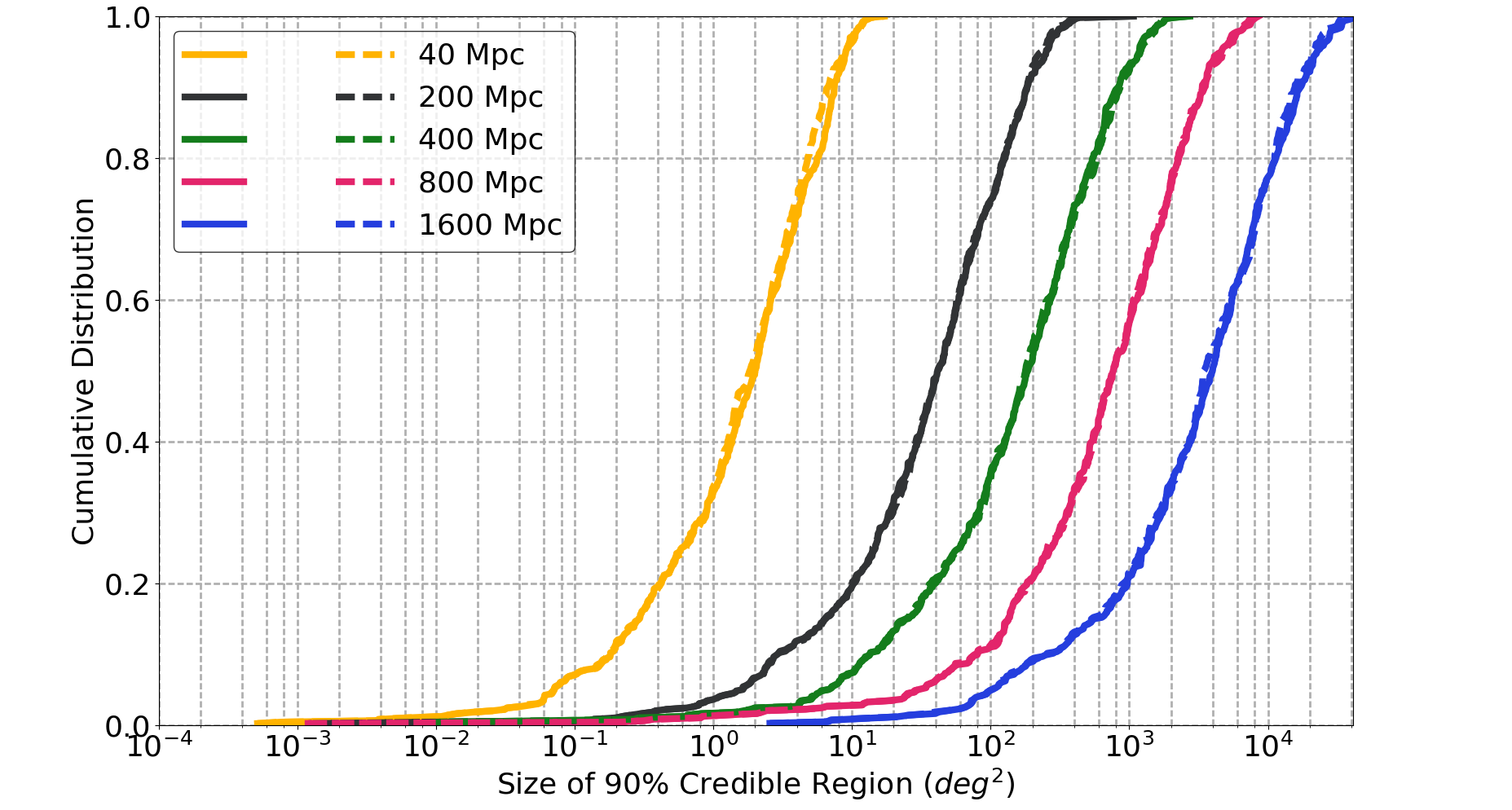}
\caption{The cumulative distribution of the size of $90\%$
credible regions for sources at fixed distances, with and without the Doppler shift effect. 
The x-axis shows the size of the $90\%$ credible region and the upper
limit of the x-axis corresponds to the size of the whole sky.
The yellow, black,  green,  red, and  blue lines represent
\ac{BNS} sources at $40$, $200$, $400$, $800$, and $1600$Mpc respectively.  The solid lines and the dashed lines show the results with and without including the Doppler shift of the waves respectively. \label{fig:withandwithout}} 
\end{figure}

To test this, we repeat the simulations for the \ac{ET} shown in Section \ref{subsec:localisation}.
While we still enable a time-dependent detector response, we fix the time delay between the center of the earth and the \ac{ET} at the beginning of the signals.
This is because turning on and off the Doppler shift should allow us to see more easily its importance. The results are shown in Figure \ref{fig:withandwithout}.
It can be seen that at all distances, the cumulative distributions are almost identical, with only marginal discrepancy.
This suggests that the Doppler effect is not important and the modulation of the detector response is the main cause of improved sky localization.

\subsection{Calibration Errors}
%
%
Previous studies have dealt with calibration errors in the context of second
generation detectors~\cite{fairhurst2009triangulation, 2012PhRvD..85f4034V,
2009PhRvD..80d2005L, 2010PhRvD..82h4020L}. We present here a brief discussion
of the impact of calibration errors on localization for third generation detectors. 
It is recognized that bias in the output of a detector can be introduced by errors 
in its calibration -- i.e., differences between the actual
response function and the measured response function of the detector.
These differences can then affect the noise and cause
amplitude errors and timing errors in the gravitational wave strain used
for analysis. Inference on the location of
the source of a \ac{GW} from the output strain can therefore be
biased.

%
Amplitude errors will affect the localization by introducing a bias in the measurement of distance, inclination and polarization angles.  With second generation detectors, these parameters cannot be measured precisely. For example, the uncertainty on distance can be $\sim40\%$ for an event with \ac{SNR} $\sim 8$~\cite{2012PhRvD..85f4034V}. Therefore, systematic errors caused by amplitude uncertainties are not expected to be dominant. However, the fractional uncertainty on distance of a \ac{GW} from \ac{BNS} is inversely proportional to \ac{SNR}. It is conceivable that when the \ac{ET} and \ac{CE} are operational, higher \ac{SNR} and the extended in-band duration will increase the accuracy with which these parameters can be determined. The amplitude error-induced bias may therefore be comparable to the uncertainty on the measurement of the parameters. Moreover, we assumed in this work that the actual value of the detector response will agree with the theoretical calculation. 
As shown in Figure \ref{fig:withandwithout}, the time evolution of the detector response is crucial for localization of \ac{BNS} mergers with 
third generation detectors. Any uncertainty in the amplitude of the waves 
or the detector response will certainly affect that.
As a result, the inference without
accounting for these errors may systematically shift the probable locations of
the source away from its true location. Amplitude errors are therefore
expected to have a larger effect in parameter estimation for third generation detectors and need to be quantified.
%

Localization can also be affected by timing errors of a signal through timing triangulation.
The accuracy with which the arrival time of a signal is determined is inversely proportional 
to the \ac{SNR} of a wave cycle at the frequencies at which the detectors are most sensitive. 
For advanced detectors, such as aLIGO and Advanced VIRGO, this happens at $\sim 100$Hz giving a timing accuracy 
$\mathcal{O}(10^{-3})$ seconds.
Timing errors (i.e., the errors intrinsic to timing when the data sample is taken) therefore would have to be comparable to a millisecond in order to be significant. However, as third generation detectors will have improved sensitivity, the \acp{SNR} for a fraction of detectable sources 
will therefore be high enough that timing error may be significant.
It is therefore necessary to quantify timing errors for third generation detectors.

\section{Conclusion}\label{sec:con}
%
%
The \ac{ET} and \ac{CE} are two currently proposed
third generation detectors. Due to the huge improvement in the sensitivity
in the frequency band below $10$Hz, the in-band durations of the gravitational waves detected from \ac{BNS} mergers will be hours or even days long. Therefore the Earth's rotation will become important, leading to several effects that become relevant for such long in-band duration signals. 
The long in-band duration allows us to observe the signal from different positions along the detector trajectory as the earth rotates.
This in turn leads to a time-dependent detector response
during the signal and also causes the wave to be Doppler
modulated. 

Using the Fisher matrix and taking the earth's rotation into consideration, 
we have estimated the localization capabilities of the \ac{ET} and \ac{CE}
individually and as a network for \ac{BNS} sources at distances equal to
$40,~200,~400,~800$ and $1600$Mpc and for a population of \ac{BNS} sources that is distributed uniformly in comoving volume. We have found that for \ac{BNS} at $40$ and $200$Mpc, the \ac{ET} alone will be able to localize most of the signals to within $100\text{deg}^2$ with $90\%$ confidence. If we assume
\ac{EM} follow up observation is achievable for \ac{BNS} whose associated
$90\%$ credible region is $\leq 100\text{deg}^2$, this means the \ac{ET} alone will be able to provide support for multi-messenger astronomy for \ac{BNS} mergers within $200$Mpc. However,
for distances $\geq 400$Mpc, localization from
the \ac{ET} alone will still be poor. This is consistent with the localization
performance for a population of \ac{BNS} distributed uniformly in comoving
volume. Of the detectable sources, only $\sim 32\%$ can be localized with $90\%$ to within a region less than the size of the whole sky.

%
Combining the \ac{ET} and \ac{CE} can dramatically boost the
performance in localization. Almost all the sources within $1600$Mpc can be
localized to within $100\text{deg}^2$ with $90\%$ confidence. In particular, the upper limit of the $90\%$ credible region for the best localized $90\%$ of the detectable sources at $40$ and $200$ Mpc has reduced by $\sim 100$ times compared to using only the \ac{ET}. Similar or greater improvements are seen for sources at greater distances.
For a population of \ac{BNS} uniformly distributed in the comoving volume, the improvement is equally impressive.  The upper limit of the $90\%$ credible region for the best localized $90\%$ of the detectable sources as derived from the Fisher matrix shrinks from an area larger than the entire sky to $\sim500\text{deg}^2$.

%
Regarding the ability to send event alerts prior to merger,
the trend is similar.  Using the \ac{ET} alone, alerts for most \acp{BNS} within
$200$Mpc can be sent a few hours prior to merger, while for \acp{BNS} at $\geq
400$Mpc, a large fraction of sources do not meet our early warning
criteria. Those which do meet the criteria do so at a time relatively close to
merger ($\mathcal{O} (10) - \mathcal{O} (10^2)$ minutes). A network with both the \ac{ET} and \ac{CE} substantially increases the number of signals at distances $\geq400$Mpc that meet the early warning criteria. This
highlights the desirability and potential of such a network for \ac{BNS} at relatively large distances.  By turning on and off the Doppler effect in the simulation, we also established that the modulation of detector responses 
during the in-band duration is the main cause for improved localization.

\section*{ACKNOWLEDGEMENTS}
We are grateful to Prof. Yanbei Chen for his help with the methodology of the paper,
and to Prof. Steve Fairhurst for his constructive comments on the paper. 
We are thankful to Dr. Xilong Fan and Teng Zhang for constructive discussions of this work. 
We are also grateful for computational resources provided by Cardiff University, 
and we are funded by an STFC grant supporting UK Involvement in the Operation of Advanced LIGO. This research is supported by The Scottish Universities Physics
Alliance and Science and Technology Facilities Council. I.S.H., C.M. and M.H. are supported by the Science and Technology Research Council (grant No. ST/L000946/1)


\end{document}